\RequirePackage{fix-cm} 
\documentclass[ twoside,openany,titlepage,numbers=noenddot,headinclude,
                footinclude=true,cleardoublepage=empty,abstractoff, 
                BCOR=5mm,paper=a4,fontsize=11pt,
                ngerman,american,%
                ]{scrreprt}

\PassOptionsToPackage{utf8}{inputenc}
	\usepackage{inputenc}

\PassOptionsToPackage{eulerchapternumbers,listings,%
					 pdfspacing,
					 subfig,beramono,eulermath,parts}{classicthesis}                                        

\newcommand{\myTitle}{Self-Supervised Beat Tracking in Musical Signals with Polyphonic Contrastive Learning\xspace}
\newcommand{\mySubtitle}{Master MVA\xspace}

\newcommand{\myName}{Dorian Desblancs\xspace}

\newcommand{\myFaculty}{Applied Mathematics\xspace}
\newcommand{\myDepartment}{Deezer Research\xspace}
\newcommand{\myUni}{École Normale Supérieure Paris-Saclay\xspace}
\newcommand{\myLocation}{Paris, France\xspace}
\newcommand{\myTime}{September 2021, revised July 2023\xspace}
\newcommand{\myVersion}{version 4.2\xspace}

\newcounter{dummy} 
\providecommand{\mLyX}{L\kern-.1667em\lower.25em\hbox{Y}\kern-.125emX\@}

	\usepackage{babel}                  

\usepackage{csquotes}
\PassOptionsToPackage{%
	backend=bibtex8,bibencoding=ascii,%
	language=auto,%
	style=numeric-comp,%
    sorting=nyt, 
    maxbibnames=10, 
    natbib=true 
}{biblatex}
    \usepackage{biblatex}

\PassOptionsToPackage{fleqn}{amsmath}       
    \usepackage{amsmath}

\PassOptionsToPackage{T1}{fontenc} 
    \usepackage{fontenc}     
\usepackage{textcomp} 
\usepackage{scrhack} 
\usepackage{xspace} 
\usepackage{mparhack} 
\usepackage{fixltx2e} 
\PassOptionsToPackage{printonlyused,smaller}{acronym} 
    \usepackage{acronym} 
    
\usepackage{tabularx} 
\usepackage{caption}
\usepackage{subfig}  

\usepackage{listings} 
\PassOptionsToPackage{pdftex,hyperfootnotes=false,pdfpagelabels}{hyperref}
    \usepackage{hyperref}  
\PassOptionsToPackage{pdftex}{graphicx}
    \usepackage{graphicx}

\hypersetup{%
    colorlinks=true, linktocpage=true, pdfstartpage=3, pdfstartview=FitV,%
    breaklinks=true, pdfpagemode=UseNone, pageanchor=true, pdfpagemode=UseOutlines,%
    plainpages=false, bookmarksnumbered, bookmarksopen=true, bookmarksopenlevel=1,%
    hypertexnames=true, pdfhighlight=/O,
    urlcolor=webbrown, linkcolor=RoyalBlue, citecolor=webgreen, 
    pdftitle={\myTitle},%
    pdfauthor={\textcopyright\ \myName, \myUni, \myFaculty},%
    pdfsubject={},%
    pdfkeywords={},%
    pdfcreator={pdfLaTeX},%
    pdfproducer={LaTeX with hyperref and classicthesis}%
}   

\makeatletter
\@ifpackageloaded{babel}%
    {%
       \addto\extrasamerican{%
                }%
       \addto\extrasngerman{%
                }%
    }{\relax}
\makeatother

\usepackage{float}
\usepackage[subfigure]{classicthesis} 

\begin{document}
\frenchspacing
\raggedbottom
\selectlanguage{american} 
\renewcommand*{\bibname}{Bibliography}
\pagenumbering{roman}
\pagestyle{plain}
\begin{titlepage}
    \begin{addmargin}[-1cm]{-3cm}
    \begin{center}
        \large  

        \hfill
        
        \vfill

        \begingroup
            \color{Maroon}\spacedallcaps{\myTitle} \\ \bigskip
        \endgroup

        \spacedlowsmallcaps{\myName}  \\ \bigskip
        \spacedlowsmallcaps{Master MVA}

        \vfill
        
        \begin{minipage}{.5\textwidth}
            \centering
            \textit{Tuteurs de stage:}\\
            \textbf{Romain Hennequin}\\
            Deezer Research\\
            \textbf{Vincent Lostanlen}\\
            CNRS, LS2N
        \end{minipage} %
        \begin{minipage}{.5\textwidth}
            \centering
            \textit{Enseignant référent:}\\
            \textbf{Roland Badeau}\\
            CNRS, Télécom ParisTech
        \end{minipage} %
        
        \vfill

        \myDepartment\\
        \myUni
        
        \vfill 
        
        \myLocation\\
        \myTime

        \vfill                      

    \end{center}  
  \end{addmargin}       
\end{titlepage}   
\thispagestyle{empty}

\hfill

\vfill

\noindent\myName: \textit{\myTitle,} \mySubtitle, 
\textcopyright\ \myTime

%
%
%
%
%

\thispagestyle{empty}
\refstepcounter{dummy}
\pdfbookmark[1]{Dedication}{Dedication}

\vspace*{3cm}

\begin{center}
    Dedicated to the loving memory of the scientist in my family, my grandfather, Popatlal Patel. \\ \smallskip
    1933\,--\,2012
\end{center}
\pdfbookmark[1]{Abstract}{Abstract}
\begingroup
\let\clearpage\relax
\let\cleardoublepage\relax
\let\cleardoublepage\relax

\chapter*{Abstract}

Annotating musical beats is a very long and tedious process. In order to combat this problem, we present a new self-supervised learning pretext task for beat tracking and downbeat estimation. This task makes use of Spleeter, an audio source separation model, to separate a song's drums from the rest of its signal. The first set of signals are used as positives, and by extension negatives, for contrastive learning pre-training. The drum-less signals, on the other hand, are used as anchors. When pre-training a fully-convolutional and recurrent model using this pretext task, an onset function is learned. In some cases, this function is found to be mapped to periodic elements in a song. We find that pre-trained models outperform randomly initialized models when a beat tracking training set is extremely small (less than 10 examples). When this is not the case, pre-training leads to a learning speed-up that causes the model to overfit to the training set. More generally, this work defines new perspectives in the realm of musical self-supervised learning. It is notably one of the first works to use audio source separation as a fundamental component of self-supervision.

\vfill

\endgroup			
\pdfbookmark[1]{Acknowledgments}{acknowledgments}


\bigskip

\begingroup
\let\clearpage\relax
\let\cleardoublepage\relax
\let\cleardoublepage\relax
\chapter*{Acknowledgments}

Before starting this masters thesis, I would like to thank some of the people that have helped me achieve this academic milestone in my young life.

First, a huge thank you to my supervisors Romain and Vincent, who gave me the opportunity to conduct research in this obscure field of machine learning. Your weekly feedback regarding my work was extremely valuable. More importantly, I really enjoyed your company on this project, and do not doubt future students of yours will too.

I would also like to thank the rest of the Deezer team, and especially the Research team. You welcomed me with open arms, and were there for me when I needed your help. On a social level, you were a lot of fun, and I hope to cross paths with you in years to come.

To my professors at the École Normale Supérieure Paris-Saclay and McGill University - thank you. I discovered a love for science and research through your lens, and truly believe the knowledge you shared with me will lead to a fulfilling career.

Finally, to my friends and family - thank you for sticking with me throughout the years. I love you all more than you could ever imagine. Mika and Zoe, thank you for being the best little siblings an older brother could ask for. Seeing us all grown-up and thriving is an amazing feeling. Mom and Dad, congratulations on 30 years of marriage! Your love and support has made everything I have achieved in my life possible. Thank you for putting up with my stubbornly stupid moments. I love you all so much.

\endgroup

\pagestyle{scrheadings}
\refstepcounter{dummy}
\pdfbookmark[1]{\contentsname}{tableofcontents}
\setcounter{tocdepth}{2} 
\setcounter{secnumdepth}{3} 
\manualmark
\markboth{\spacedlowsmallcaps{\contentsname}}{\spacedlowsmallcaps{\contentsname}}
\tableofcontents 
\automark[section]{chapter}
\renewcommand{\chaptermark}[1]{\markboth{\spacedlowsmallcaps{#1}}{\spacedlowsmallcaps{#1}}}
\renewcommand{\sectionmark}[1]{\markright{\thesection\enspace\spacedlowsmallcaps{#1}}}
\clearpage

\begingroup 
    \let\clearpage\relax
    \let\cleardoublepage\relax
    \let\cleardoublepage\relax
    \refstepcounter{dummy}
    \pdfbookmark[1]{\listfigurename}{lof}
    \listoffigures

    \vspace{8ex}

    \refstepcounter{dummy}
    \pdfbookmark[1]{\listtablename}{lot}
    \listoftables
        
    \vspace{8ex}
    

    \vspace{8ex}
       
\endgroup

\pagenumbering{arabic}
\chapter{Introduction}\label{ch:introduction}

Can knowledge acquired from performing one task be transferred over to another? 
For humans, the answer is most likely yes \citep{valentin2007determining} \citep{soltani2019adaptive} \citep{gerraty2014transfer}. Over the past few decades, neuroscientists have been trying to figure out how synapses in the brain adapt to being exposed to new stimuli.
The field of goal-directed learning \citep{valentin2007determining} has notably helped us uncover how neural pathways are modulated when an individual is exposed to novel situations. It has shown that humans are constantly adapting past knowledge to new tasks to learn more efficiently \citep{soltani2019adaptive} \citep{gerraty2014transfer}.



In the field of artificial intelligence, recent works have suggested that deep neural networks are also capable of transferring the knowledge they gain when being trained on one task to learning about another task \citep{cramer2019look} \citep{zhao2020contrastive}. Networks that perform well on large datasets such as ImageNet \citep{deng2009imagenet} have been successfully fine-tuned to more precise tasks such as dog identification \citep{huh2016makes}. This is most likely due to the fact that models trained on the ImageNet dataset already possess an understanding of the world we live in (the ImageNet dataset contains 200 classes that span a wide variety of objects). Transferring this understanding to more precise tasks has proven to be extremely beneficial, whether it be for speeding up learning or increasing a model's performance. 

Knowledge transfer is often done in one of two ways. 
First, a network can be frozen and fine-tuned to a new task by merely training the topmost layers. 
Frozen layers output a representation that is then adapted to a new task by training an extra layer or two.
Second, a network can be fine-tuned to a new task by being fully re-trained. 
In this case, however, a very small learning rate is used, so as to avoid changing the network too drastically in the first few epochs.

The success of transfer learning has inspired researchers greatly. Although machine learning has come a long way, the task of labelling thousands, if not millions, of videos, sounds, or other samples is extremely tedious and costly. What if neural networks could be trained on extremely large datasets that require minimal labelling? Two popular machine learning sub-fields aim to do so: unsupervised learning and self-supervised learning. Both of these forms of learning allow a model to learn without hand-crafted labels. In the former, the data used to train a model is fully untagged. The model then aims to learn a representation of its inputs. Popular unsupervised learning networks include autoencoders \citep{kramer1991nonlinear}, which attempt to efficiently encode and regenerate an input sequence, and Generative Adversarial Networks (GAN) \citep{goodfellow2014generative}, which generate new data that has the same properties as the training set. 

Self-supervised learning, on the other hand, is a form of learning in which data are labelled in a computer-generated fashion. Data samples are not labelled using human knowledge. Instead, they are labelled automatically. In this form of learning, we distinguish two sets of tasks. The first are referred to as pretext tasks. In these, models have to solve a simple problem, such as a classification problem, using the computer-generated labels. Models trained for these tasks often process large amounts of training data, so as to learn a desired representation of their inputs. On the other hand, the tasks on which a model is fine-tuned are referred to as downstream tasks. The datasets associated with these tasks are often smaller and human-labelled. 

\begin{figure}[h]
    \centering
    \includegraphics[width=1\textwidth]{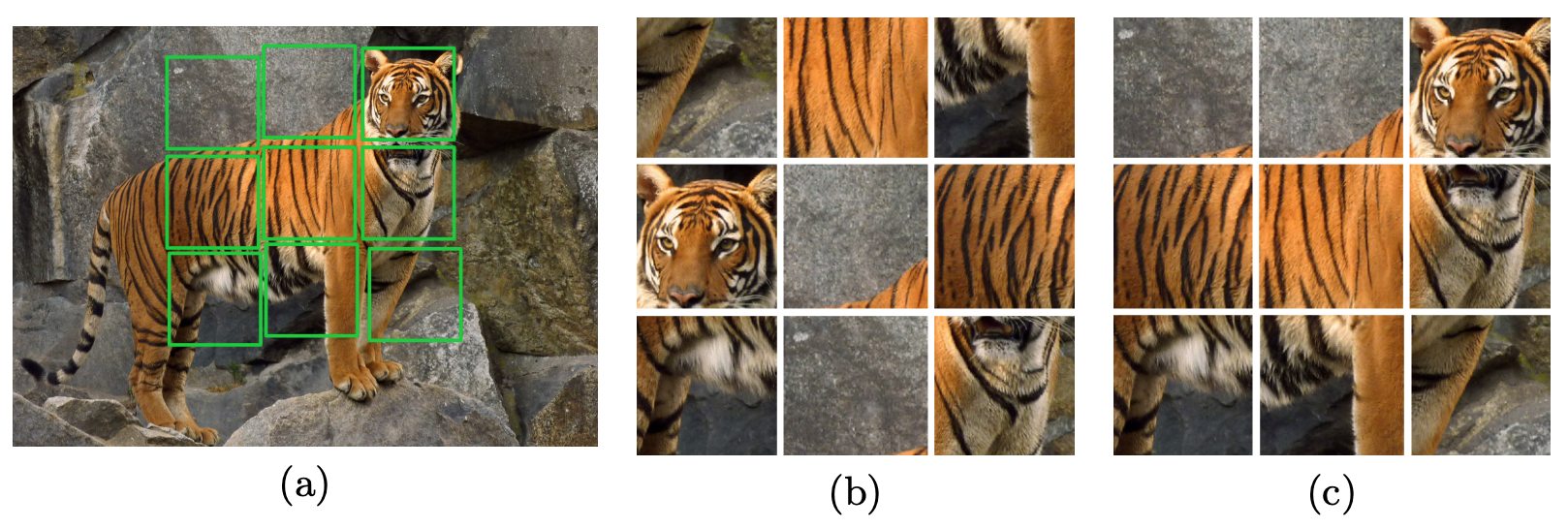}
    \caption[Self-supervised Learning via Solving Jigsaw Puzzles of natural Images \citep{noroozi2016unsupervised}]{Self-supervised Learning via Solving Jigsaw Puzzles of natural Images \citep{noroozi2016unsupervised}. (a) displays the segmentaion of an image into multiple puzzle pieces. (b) displays the shuffled pieces of the puzzle and is used as input to the model. Finally, (c) shows the desired model output; these are the unshuffled pieces of the puzzle.}
    \label{fig:jigsaw}
\end{figure}

In computer vision, for example, one of the first pretext tasks ever introduced can be found in \citep{noroozi2016unsupervised}. In their paper, the authors demonstrate that solving jigsaw puzzles of natural images allows a neural network to learn general representations of the world (i.e. learning features that can then easily be adapted to a wide range of vision tasks). Figure \ref{fig:jigsaw} outlines their methodology. The neural network's weights can then be fine-tuned to a variety of downstream tasks. In each of these, the performance achieved by pre-trained models is greater than that of their purely supervised counterpart. More importantly, self-supervised pre-training is shown to bridge the gap with ImageNet pre-training, while having fully eliminated the cost of human labelling. 

The work we present in this thesis is directly linked to self-super-vised learning, and more specifically one of its main sub-components: contrastive learning. In this form of learning, networks learn how to distinguish sample pairs (two distinct segments of a same image for example) from other samples (segments of other images for example).
Contrastive learning was popularized in 2020 by Hinton et al. \citep{chen2020simple}, and is gaining traction in the field of deep learning. We will explore the paradigm more deeply later in this report. Our work focuses on contrastive learning applied to the field of Music Information Retrieval (MIR). 

MIR is a broad and interdisciplinary field dedicated to the "understanding, processing, and generation of music \citep{peeters2021deep}." We understand music by assigning tags to it, estimating its pitch, or detecting its mood. We process music by separating a song into its different sources. Finally, we can generate music by transferring features or characteristics of one song to another. The above are all examples of common problems being solved in the field of MIR. As in many fields, deep learning has become the most widespread approach in many tasks. This is notably the case in musical source separation, where Spleeter \citep{spleeter2020}, a deep learning network, has become a standard tool for separating songs into their various stems (drums, vocals...). We refer to \citep{peeters2021deep} for an overview of how deep learning is applied to audio and music today.

Two of the most popular tasks in MIR are associated to the beat of a song. The beat is often described as the rhythm a listener taps his foot to when listening to a piece of music. Beats are also referred to as pulses, and are a fundamental time unit for understanding music. We refer to beat tracking as the task of estimating the times at which a beat occurs throughout a song. We refer to downbeat estimation as the task of estimating the first beat of each bar. Recent works in the field have focused on fully-convolutional and/or recurrent model architectures \citep{matthewdavies2019temporal} \citep{bock2016joint}. In these works, the output of the model is an activation function whose values are closer to 1 when a beat or downbeat occurs, and closer to 0 when this is not the case.\footnote{Note that a Dynamic Bayesian Network (DBN) is then used to "pick" the beats from the activation function. We will come back to this later in the report.} The inputs are usually log mel-spectrograms.

Annotating a song's beats and downbeats is however an extremely lengthy and time-consuming process.\footnote{One can read more about the beat annotation process \href{https://musicinformationretrieval.wordpress.com/2017/04/25/audio-beat-tracking-human-annotation-strategies/}{here}.} In this research project, we use contrastive learning and musical source separation to create a novel pretext task for beat tracking and downbeat estimation. This pretext task can be formulated as trying to match a song's drums to the rest of the signal (i.e.: a mixture of its vocals, bass, and other components). By doing so, we postulate that our model gains knowledge about what constitutes a beat, and how to identify one from a short spectrogram excerpt. The model we design is trained on 40+ hours of music and fine-tuned on popular beat tracking and downbeat estimation datasets. Our contributions to the fields of MIR and deep learning can be summarized as follows.

We introduce a new pre-training method for tasks related to musical beat identification. More generally, we use audio source separation as a fundamental component of self-supervision, and believe that the use of such audio processing could be beneficial for future musical self-supervised pretext tasks. The result of this pre-training is a model that automatically learns an onset function. Our pre-training method enables faster and more efficient training on popular beat tracking and downbeat estimation datasets.

In chapter 2, we briefly review some key audio time-frequency representations. Understanding these is essential for comprehending both our work and the works that inspired us.
Chapter 3 introduces a number of key works related to our experiments. It notably explores how self-supervised learning and contrastive learning have been used in audio and MIR. It also delves into the state-of-the-art methods used for beat tracking and downbeat estimation.
In chapter 4, we explore our methodology in depth. We first detail our network's design and how we process our data. We then outline the algorithms used for our pretext and downstream tasks.
Chapter 5 displays the results we obtain on a variety of experiments, such as pure beat tracking, joint beat and downbeat estimation, and cross-dataset generalization. Note that in each case, the effect of pre-training is analyzed thoroughly. 
Finally, in chapter 6, we discuss the limitations of our work. We also present new research directions that involve both source separation and self-supervised learning. A short conclusion ensues.



\chapter{Background Information}\label{ch:introduction}

Before delving into the related works section, let us briefly introduce some of the key concepts behind audio representations. In the field of deep learning, many different audio representations have been used as inputs to a network, including waveform representations of a signal. Time-frequency representations are however much more common. Over the years, many alterations of the basic spectrogram have been introduced to better understand audio. We will introduce some of the most popular ones. We will start with a simple review of spectrograms before making our way to the Variable-Q Transform (VQT), which was used in our work. Note that a more in-depth explanation of these concepts can be found in \citep{peeters2021deep} and \citep{cwitkowitz2019end}. This chapter is targeted to audiences that may not have a background in signal processing.


At its core, an audio signal $x(t)$ describes the evolution of a sound wave $x$'s pressure over time $t$. When a signal is digitized, the time dimension is dicretized across samples $m$. The result of this process is a discrete sequence $x(m)$. We refer to the sampling rate as the number of samples per second. The Discrete Fourier Transform (DFT) is used to represent a discrete, non-periodic signal over $N$ frequencies. For discrete frequencies $k \in [0, N-1]$, we can calculate the DFT using:

\begin{equation}
    X(k) = \sum_{m=0}^{N-1}x(m) e^{-j2\pi \frac{k}{N}m}
\end{equation}



\begin{figure}[h]
    \centering
    \includegraphics[width=.85\textwidth]{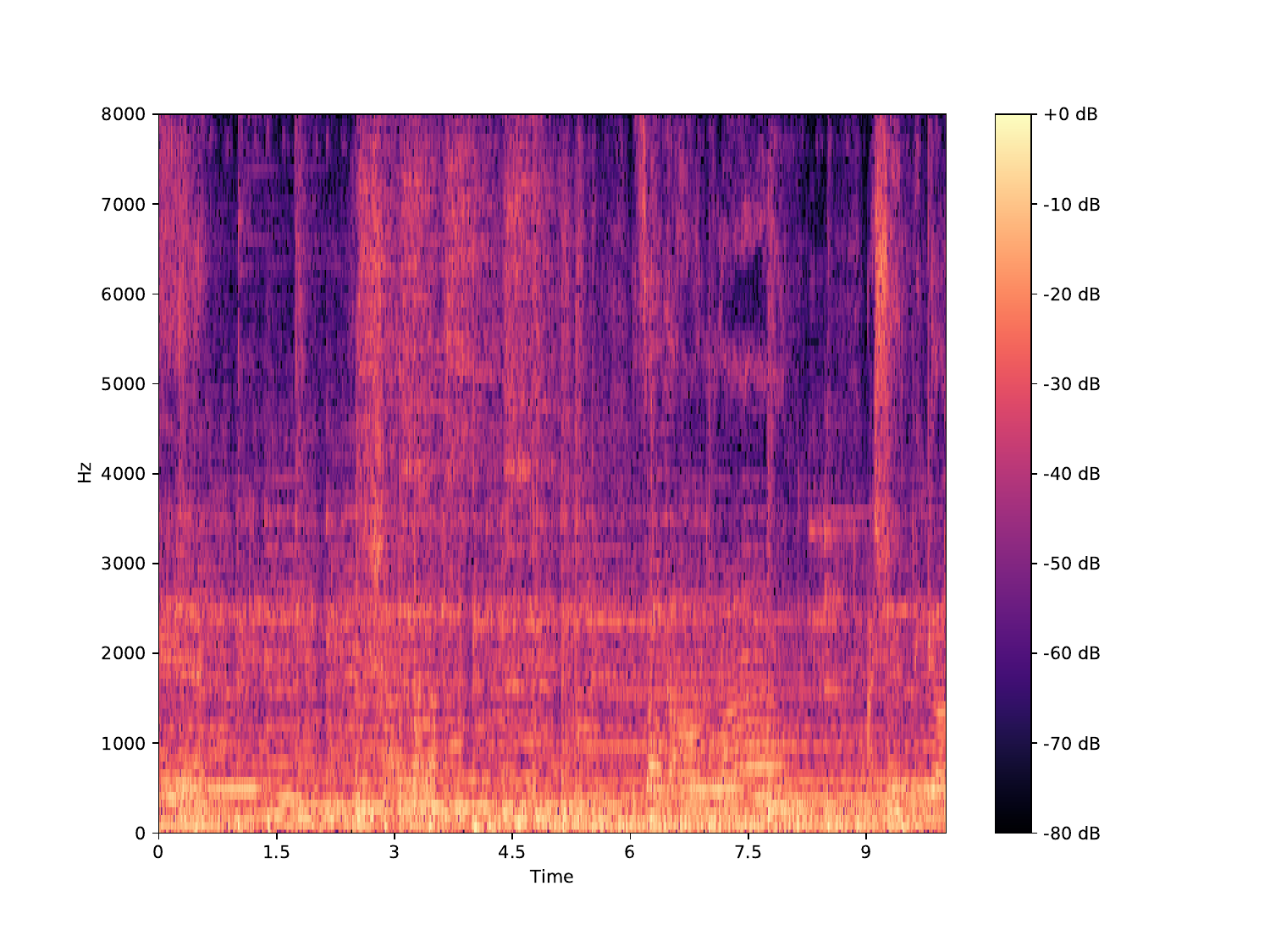}
    \caption{Spectrogram Representation}
    \label{fig:fft}
\end{figure}

Since audio signals vary over time, the DFT is usually calculated over consecutive frames. 
The resulting complex matrix depicts a signal's frequency evolution over time. 
The amplitude of each value is then taken to obtain a spectrogram representation of a signal. 
This is the most common audio input representation. 
Figure \ref{fig:fft} displays the spectrogram representation (in decibels) of a musical signal containing drums.\footnote{Librosa \citep{mcfee2015librosa} was used in order to generate the spectrograms in this section. Note they all have the a total of $96$ frequency bins.} 
Spectrogram bins are distributed evenly across the frequency spectrum.
Human hearing, on the other hand, works on a scale that is close to a logarithmic scale.

\begin{figure}[h]
    \centering
    \includegraphics[width=.85\textwidth]{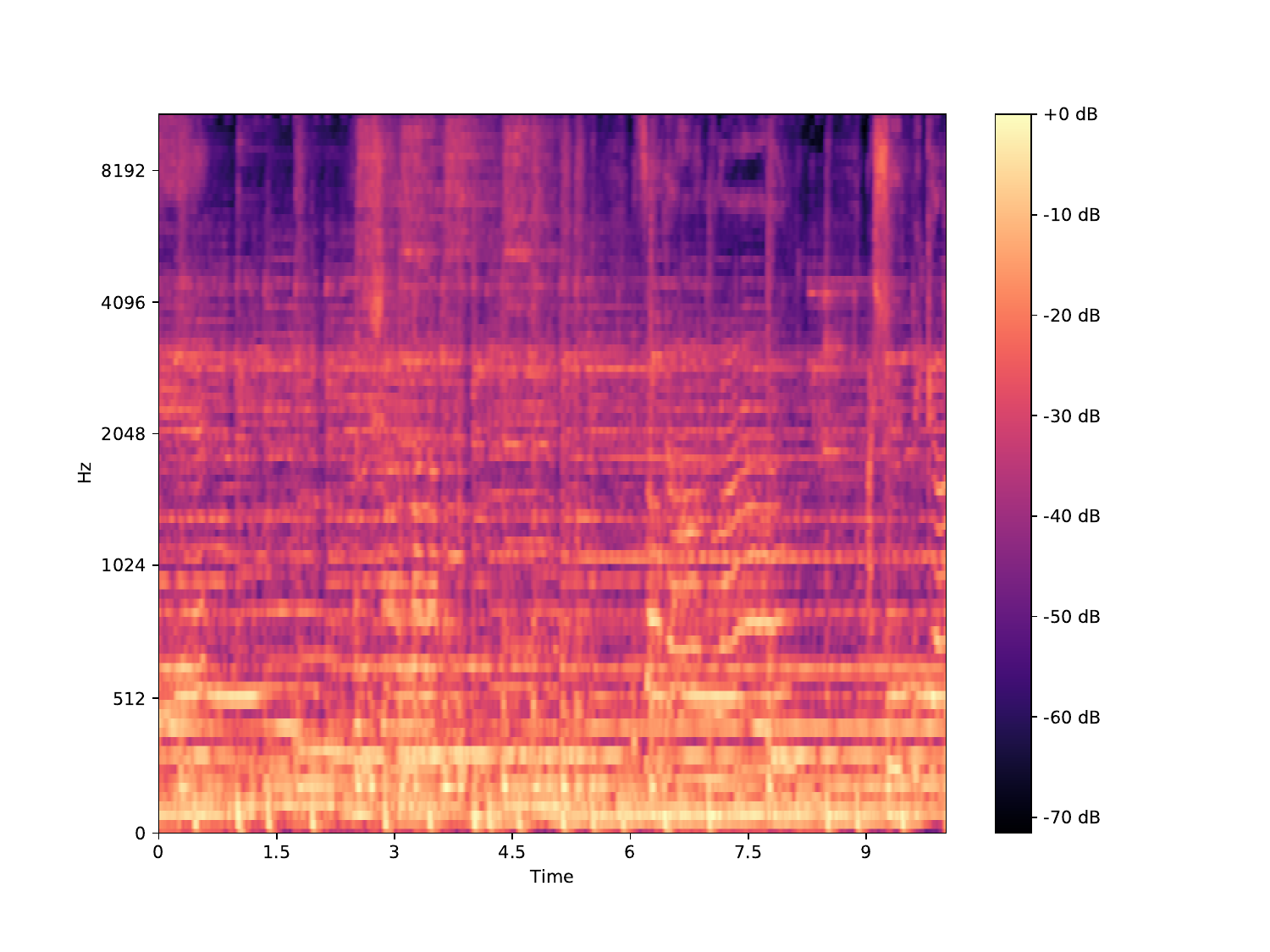}
    \caption{Mel-spectrogram Representation}
    \label{fig:mel}
\end{figure}

This has led to the adoption of numerous psycho-acoustically motivated audio representations. These are usually more compact (i.e. contain less frequency bins).\footnote{This is the main advantage of mel-spectrograms. A sound can be represented more compactly without losing much of its perceptual information.} For speech-related tasks, mel-spectrograms are extremely popular. The mel scale is a quasi-logarithmic \citep{mcfee2015librosa} function of acoustic frequencies. It is designed such that perceptually-spaced pitch intervals, such as octaves, are evenly distanced across frequency bins. Figure \ref{fig:mel} displays the mel-spectrogram of the same audio signal as Figure \ref{fig:fft}. Mel-spectrograms also tend to require less frequency bins than regular spectrograms (i.e. the frequency precision loss between a regularly-space frequency scale and a mel scale is compensated by the perceptual quality of the mel-scale).

For music-related tasks, the  Constant-Q Transform (CQT) representation has grown in popularity. The CQT is based on wavelet analysis \citep{brown1991calculation} adapted for musical signals. The centre frequencies of each bin are associated with musical note frequencies, which allows one to distinguish the musical pitches present in a signal. Hence, a pitch shift merely results in a vertical shift on the CQT representation. The representation has a logarithmic frequency scale.  

\begin{figure}[h]
    \centering
    \includegraphics[width=.85\textwidth]{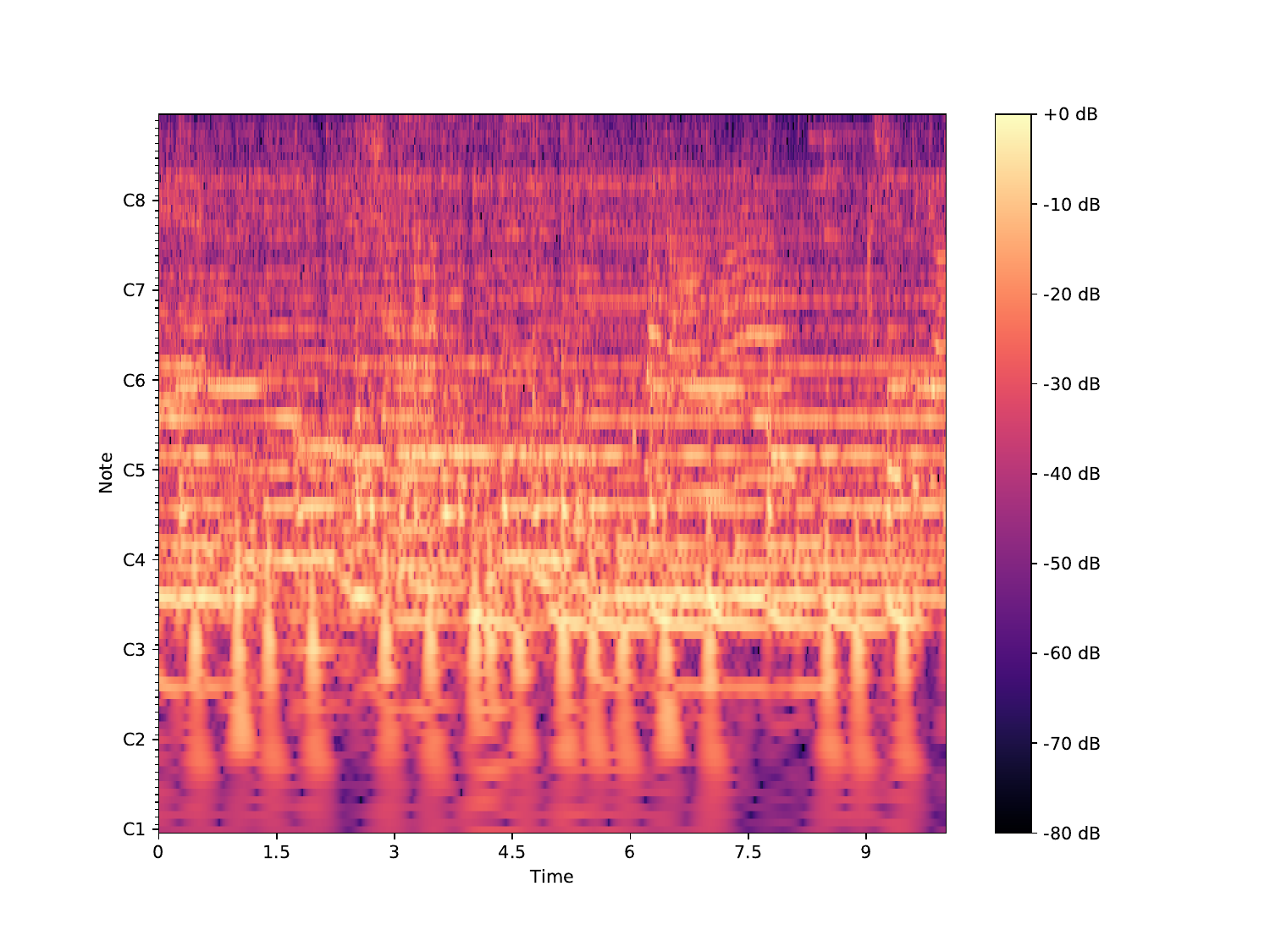}
    \caption{Constant-Q Transform Representation}
    \label{fig:cqt}
\end{figure}

\begin{figure}[h]
    \centering
    \includegraphics[width=.85\textwidth]{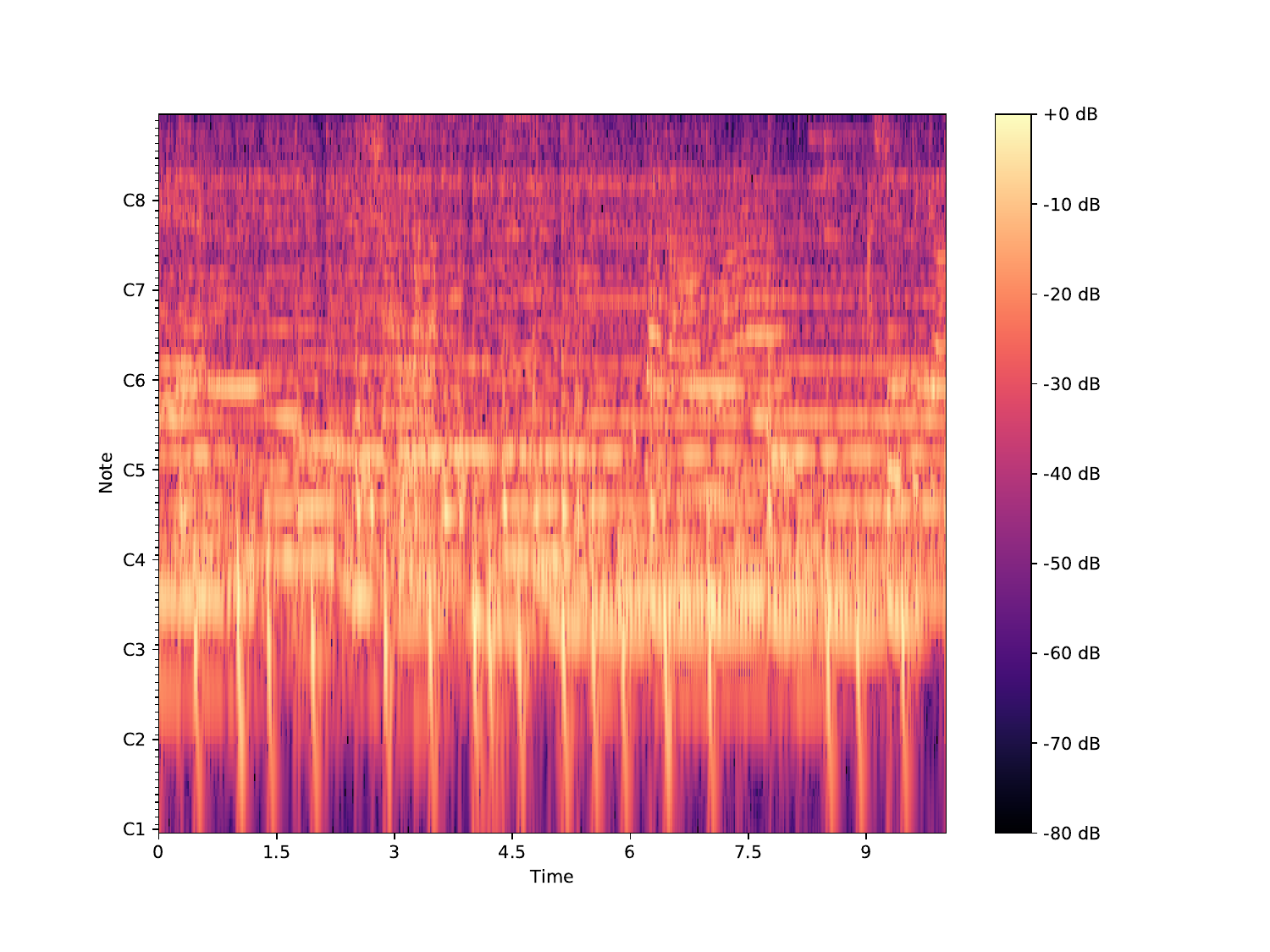}
    \caption{VQT Representation}
    \label{fig:vqt}
\end{figure}

One downside of the CQT, however, is that its time resolution at lower frequencies is quite poor. For songs that contain percussive sounds such as drums, this can be problematic. The VQT was therefore introduced to combat this problem. Frequency centres remain unchanged, but the time resolution at lower frequencies is much higher. Figures \ref{fig:cqt} and \ref{fig:vqt} display a CQT and VQT representation of our original audio signal. Notice how much more precise drum sounds appear on Figure \ref{fig:vqt}. This is especially the case for the kick drum, whose frequencies are partly concentrated between 32 and 130 Hz. 

In many cases, the input representation used for a deep learning model has a direct influence on the model's performance on a certain task \citep{huzaifah2017comparison}. In our work, we opted to use the VQT representation as an input to our model. The resolution present at lower frequencies was especially important for our work, due to the importance of accurately displaying drums for our pretext task. We will come back to the specifics of our VQT representation in Chapter 4. For now, the concepts introduced in this chapter should allow the reader to comprehend the Related Works section of this thesis. We emphasize that \citep{peeters2021deep} and \citep{cwitkowitz2019end} are more detailed resources for comprehending the specifics of each time-frequency representation introduced in this chapter.

\chapter{Related Work}\label{ch:examples}

In this chapter, we will explore the works that inspired this thesis. We will start with the field of audio representation learning. This broad field is concerned with teaching neural networks how to recognize patterns in sound. Recently, many works have focused on self-supervised learning and contrastive learning for learning better representations. We will present these. We will also present earlier papers in the fields of audio-visual correspondence and metric learning that had an impact on the field. Note that many works in the field of computer vision inspired the papers presented in this section \citep{misra2016shuffle} \citep{gidaris2018unsupervised}. 

We will then briefly introduce some of the important works in the field of audio source separation, before focusing on beat tracking and downbeat estimation. For the latter, both classic and modern methods will be presented. We distinguish these as follows: modern methods make use of neural networks for beat estimation, whereas classical methods do not. Although this is a gross simplification, we do so for clarity purposes.

\section{Audio Representation Learning}

\subsection{Audio-Visual Correspondence}

\begin{figure}[H]
    \centering
    \includegraphics[width=.7\textwidth]{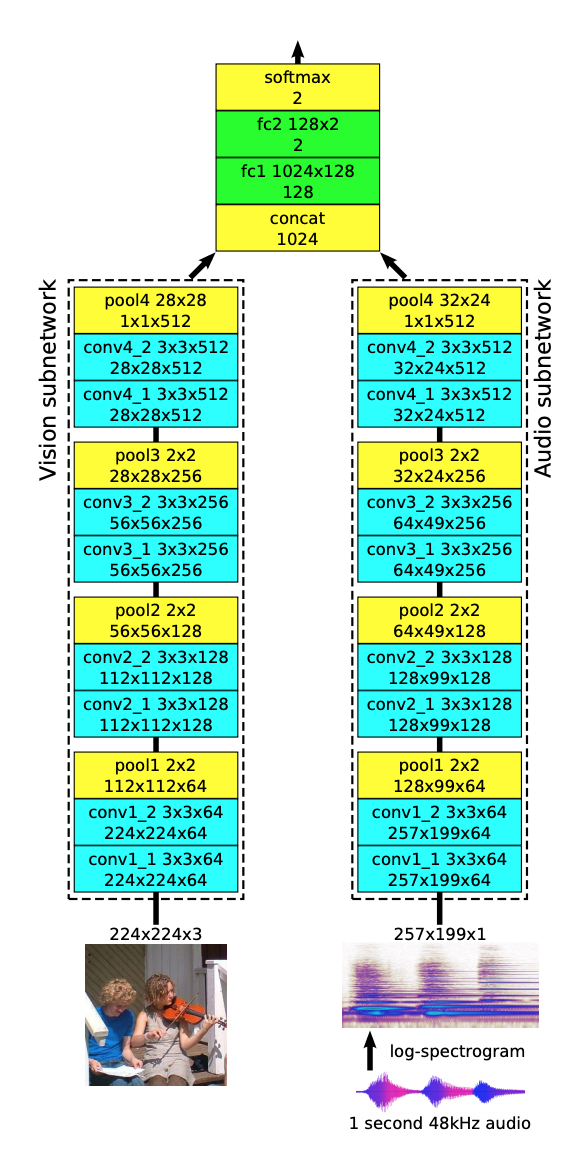}
    \caption{Look, Listen, and Learn Audio-Visual Correspondence Architecture \citep{arandjelovic2017look}}
    \label{fig:l3}
\end{figure}

One of the first papers to explore the field of audio representation learning was \textit{Look, Listen, and Learn} \citep{arandjelovic2017look}. In their paper, Zisserman et al. try to match a sound to an image. This is done by extracting sounds and images from video frames. From there, two neural networks are used to determine whether a sound corresponds to this image. A vision subnetwork studies the image input, while an audio subnetwork studies the log-spectrogram of the audio input. The network architecture for \textit{Look, Listen, and Learn} \citep{arandjelovic2017look} can be found in Figure \ref{fig:l3}. The outputs of each subnetwork are then fused together to determine whether the image and sound come from the same video. This work is significant for many reasons. In the realm of sound, extracting the audio subnetwork and fine-tuning it to a downstream task leads to much better performance on the downstream task. Why? Because this pretext task allows the network to distinguish sounds by matching them to an image. Note that the work in \citep{arandjelovic2017look} has been enhanced by using log mel-spectrograms as an input to the audio subnetwork \citep{cramer2019look}.

\begin{figure}[H]

\centering

\subfloat[Audio Source Separation Pipeline]{\includegraphics[width=1\textwidth]{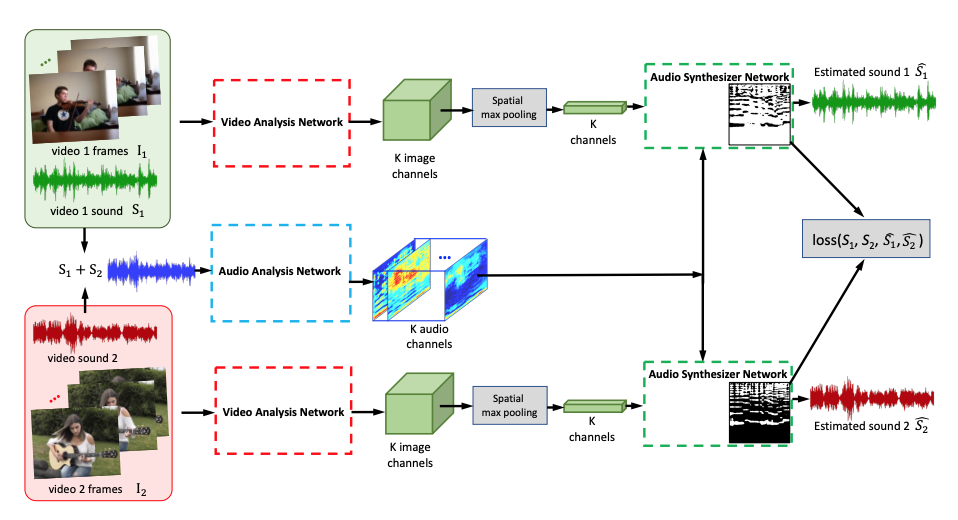}}
\newline
\subfloat[Sample Sound Localization Results]{\includegraphics[width=1\textwidth]{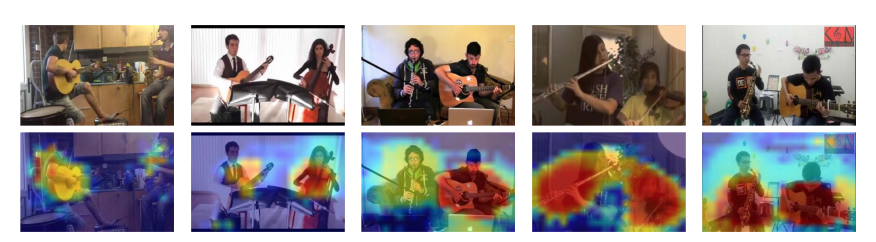}}

\caption{\textit{The Sound of Pixels} \citep{zhao2018sound} Task Diagram and Results}

\label{fig:sop}

\end{figure}

In 2018, Zhao et al. \citep{zhao2018sound} used a similar idea to perform audio source separation. In their work, image frames from two separate videos are used as input to a visual model. The sounds corresponding to these image frames are then mixed and input to an audio network. The goal of their pretext task is to separate the mixture into two audio snippets, corresponding to the separated sounds of each video. Their method achieves extremely good results in the realm of sound separation. The networks are notably found to match sounds to specific objects in an image. Figure \ref{fig:sop} illustrates their methodology and results.

Although early works in the field of audio representation learning focus on audio-visual correspondences, a flurry of papers in the realm of pure audio representation learning were published around the same time. These notably make use of triplet-based metric learning, and will be presented in the next section.

\subsection{Triplet-Based Metric Learning}

At its most basic level, triplet-based metric learning is used to train a network to distinguish pairs of images, time series, or other data types. Assuming a triplet $t = (x_a, x_p, x_n)$, we define:

\begin{itemize}
    \item the anchor sample $x_a$
    \item the positive sample $x_p$
    \item the negative sample $x_n$
\end{itemize}

The goal of a triplet learning task is then to distinguish the anchor and positive from the negative. The anchor and positive are usually either two samples from the same class, or augmented versions of each other. Popular loss functions include margin loss, which is defined as:

\begin{equation}
    L(t)=max[0, D(f(x_a), f(x_n))-D(f(x_a), f(x_p))+\delta]
\end{equation}

where $D$ is a distance metric\footnote{Popular margin loss distance metrics are cosine similarity and euclidean distance. The margin value is usually adapted to the chosen distance function. For example, in the case of cosine similarity, $\delta$ is between 0 and 1. As $D(f(x_a), f(x_p))$ gets closer to 1, and as $D(f(x_a), f(x_n))$ gets closer to -1, the loss decreases.} and $\delta$ is referred to as a margin value \citep{weinberger2009distance}. Note that each anchor/positive pair can be compared to multiple negatives at a time, in which case the above formula is used for every negative. The values returned are then summed or averaged.

Triplet-based learning is used in three works that greatly inspired us. The first was published in 2018 \citep{jansen2018unsupervised}. In this paper, Jansen et al. sample random snippets of audio from each data point in the AudioSet \citep{gemmeke2017audio}. These are the anchors. Each anchor is then augmented using a series of signal transformations such as Gaussian noise addition and time-frequency translation to create positive samples. A simple ResNet \citep{he2016deep} model is then trained using margin loss with a cosine similarity distance metric. The model is fine-tuned on two downstream tasks: query-by-example and sound classification. The fine-tuned models prove to perform very closely to their fully-supervised counterparts using extremely little data.

Lee et al. \citep{lee2020metric} use a similar approach to learn musical representations. In their work, sample pairs are snippets of songs that have the same class (i.e. same genre). Triplet learning is found to sucessfully pre-train their model on two musical tasks: similarity-based song retrieval and auto-tagging.

\begin{figure}[h]
    \centering
    \includegraphics[width=1\textwidth]{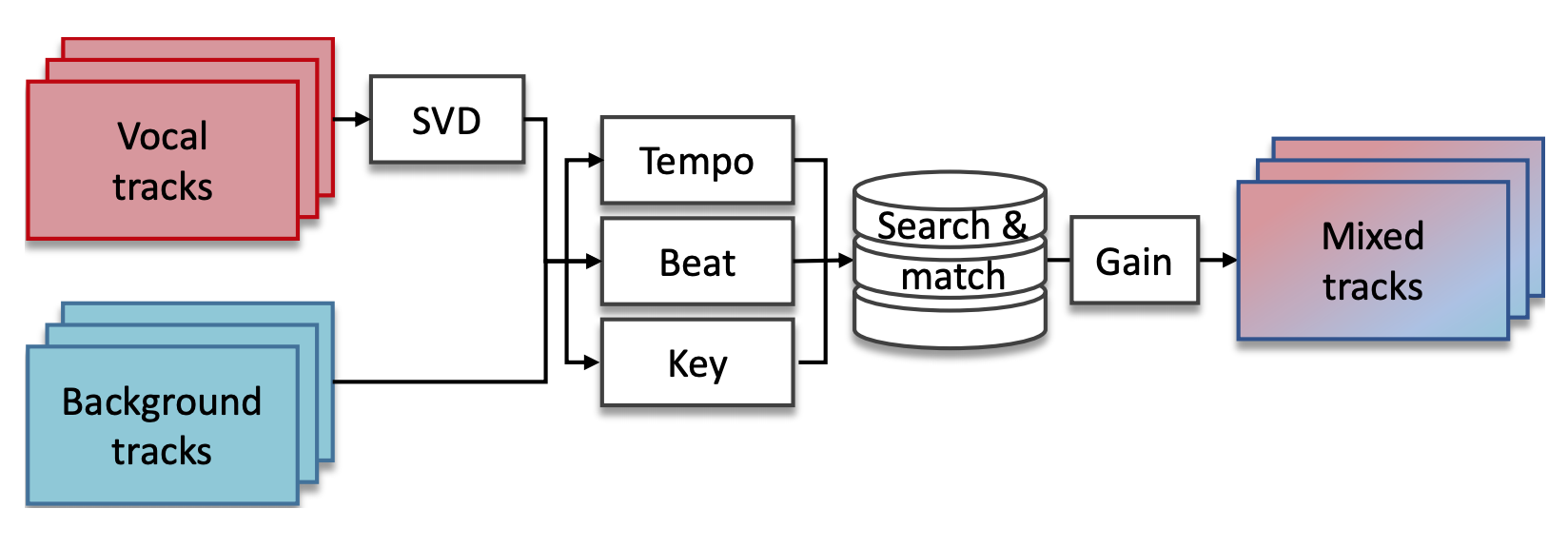}
    \caption{Vocal and Background Track Mashup Pipeline \citep{lee2019learning}}
    \label{fig:mashup}
\end{figure}

Finally, the work on triplet learning that inspired our experiments the most was published by Lee et al. in 2019 \citep{lee2019learning}. Their work relies on mashing up vocals and background tracks that have similar tempo, beat, and key. From there, triplets are generated using tracks that contain the same vocals, but a different background track. They use a margin loss. The goal of their pretext task is to train a model to recognize a vocal from the same singer amidst a different background track.\footnote{Note that three settings are used in this triplet learning task. A MONO setting, where all inputs are monophonic (i.e. only contain vocals), a MIXED setting, where all inputs are mashups, and a CROSS setting, where anchors are monophonic and positives are mashups.} The outputs of pre-trained network are then adapted for two musical tasks: singer identification and query-by-singer. In both tasks, Lee et al. achieve extremely high accuracy. The audio mashup pipeline used in \citep{lee2019learning} is illustrated in Figure \ref{fig:mashup}.

This work greatly inspired us because of its ingenious use of musical stems. By distinguishing vocals that come from the same song/ artist from their background songs, a neural network can learn accurate voice embeddings that can be used for a variety of musical tasks. We separate drum stems from the rest of our songs in order to learn a more rhythmic representation of music for tasks such as beat tracking and downbeat estimation. 

\subsection{Self-Supervised Learning}

As mentioned previously, self-supervised learning involves a pretext task, used to pre-train a neural network using lots of data, and a downstream task that is often quite precise and limited in its data. When it comes to designing a pretext task, there exist a wide range of options. In some cases, the tasks are quite general, and are aimed towards learning an audio representation that can span numerous downstream tasks. In other cases, however, the pretext task is geared towards a specific downstream task.

\begin{figure}[H]
    \centering
    \includegraphics[width=1\textwidth]{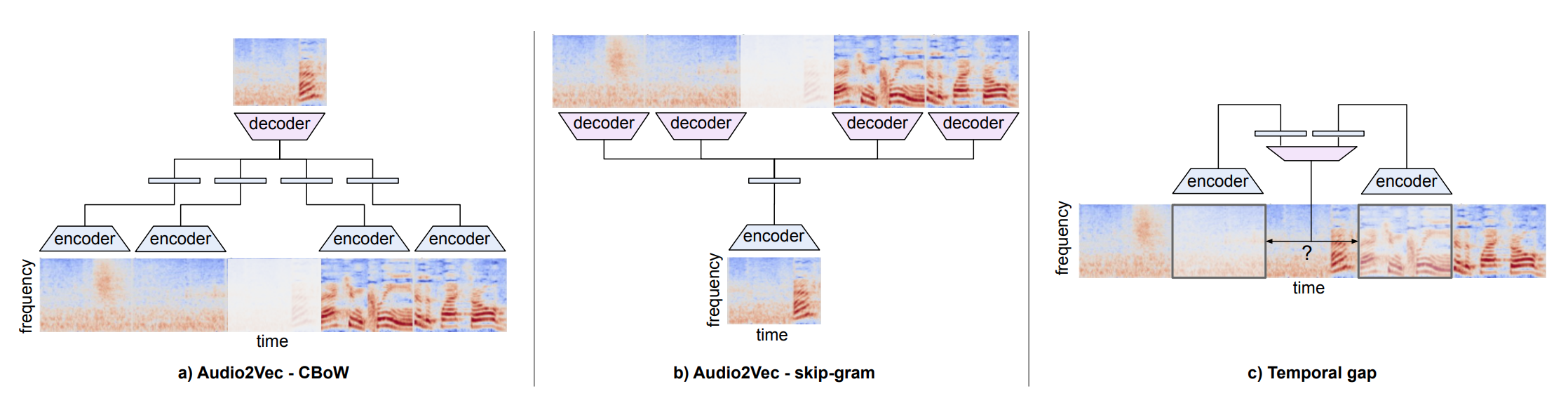}
    \caption[Overview of the Self-Supervised Learning Tasks Introduced by Tagliasacchi et al. \citep{tagliasacchi2019self} \citep{tagliasacchi2020pre}.]{Overview of the Self-Supervised Learning Tasks Introduced by Tagliasacchi et al. \citep{tagliasacchi2019self} \citep{tagliasacchi2020pre}. Note that in the tasks that contain multiple inputs, all inputs are passed to the same encoder. Its outputs are then concatenated and either passed to a decoder (in CBoW) or a regular feedforward network (in temporal gap) for further processing.}
    \label{fig:taglia}
\end{figure}

Let us first explore the former. Some of the most interesting pretext tasks used in audio are introduced by Tagliasacchi et al. \citep{tagliasacchi2020pre} \citep{tagliasacchi2019self}. These are partially inspired by the famous word2vec word vectorial representations \citep{mikolov2013efficient}. In their audio2vec tasks, an autoencoder is used to reconstruct missing pieces of a log mel-spectrogram. This task comes in two forms. In its CBoW variant, the autoencoder must reconstruct a central piece of the input spectrogram. In its skip-gram version, the autoencoder must reconstruct the pieces around the central section of a spectrogram. They also introduce a third task coined temporal gap, in which a model must estimate the duration between two sections of a spectrogram. All three tasks are illustrated in Figure \ref{fig:taglia}.

The encoders from the pretext tasks are then isolated, frozen, and fine-tuned using extra linear layers for tasks such as speech recognition and urban sound classification. Merely training the linear layers on top of the learned representation leads to performances that are almost on par with fully-supervised, state-of-the-art results.

Carr et al. use similar ideas in \citep{carr2021self}. In their case, they split their input log spectrograms into a nine-piece jigsaw puzzle. They then use an autoencoder network to predict the correct permutation ordering. The encoder network is then extracted and fine-tuned using all layers (i.e. the encoder is not frozen). The performance they obtain surpasses end-to-end fully-supervised training on instrument family detection, instrument labelling, and pitch estimation tasks.

Finally\footnote{A last, non-musical, paper worth mentioning is \citep{ryan2020using}. The authors use self-supervised learning applied to bird songs for downstream industrial audio classification.}, in the realm of music, Wu et al. \citep{wu2021multi} use an encoder network to predict input song snippets' classic music features such as MFCCs, Tempograms, and Chromas. The encoder is then extracted and trained on downstream datasets such as the FMA genre \citep{defferrard2016fma} by adding an MLP on top of the network. The encoder is trained in a fully-supervised fashion (with random initializations), in a frozen fashion (pre-trained layers frozen), and a fine-tuning fashion (pre-trained layers also trained). The pretext task learned representations allowed the network to achieve results on par with end-to-end supervised training in the frozen context, and superior to end-to-end supervised training in the fine-tuning context.

\begin{figure}[h]
    \centering
    \includegraphics[width=1\textwidth]{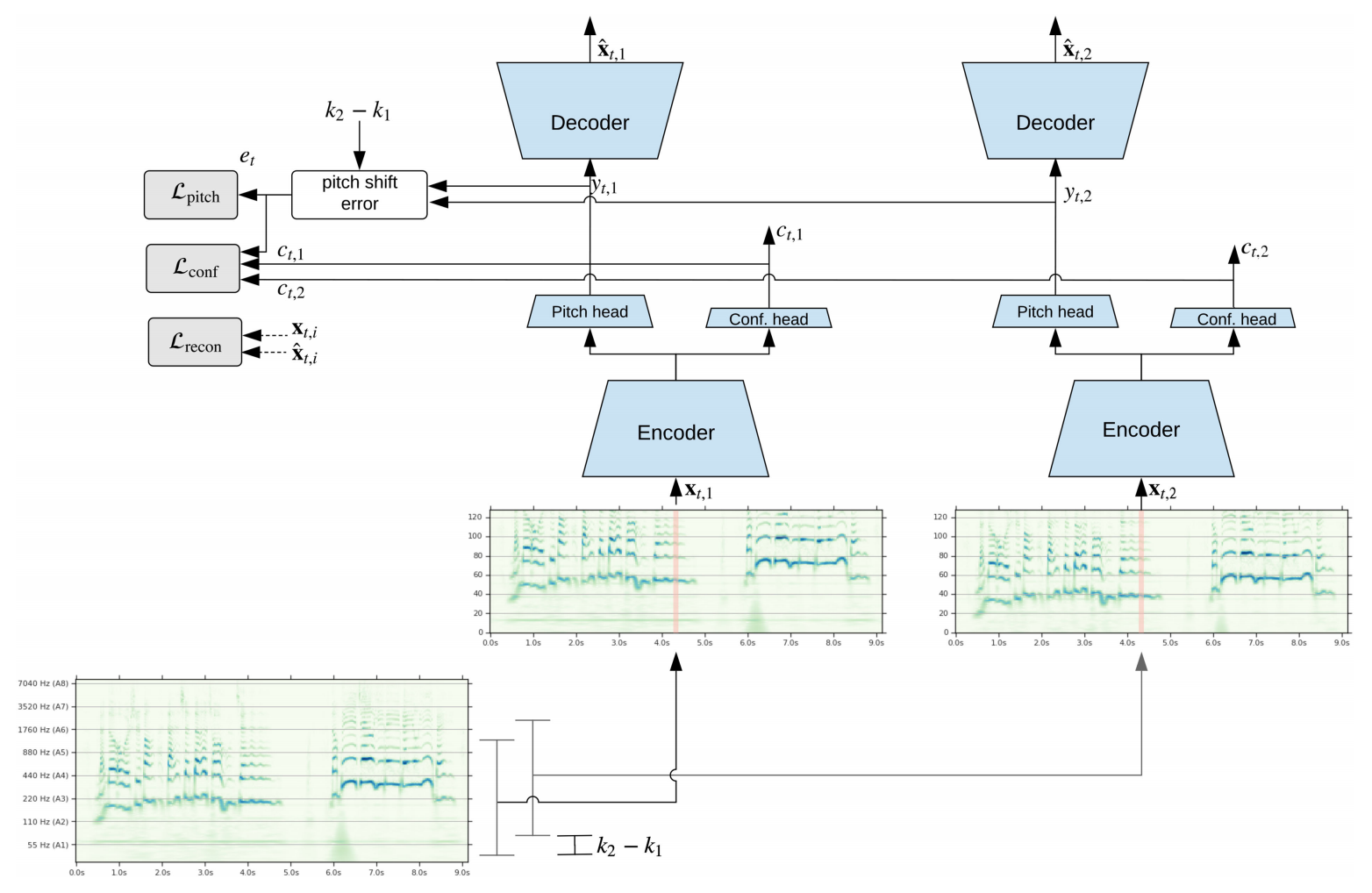}
    \caption[SPICE \citep{gfeller2020spice} Pretext Task Overview]{SPICE \citep{gfeller2020spice} Pretext Task Overview.}
    \label{fig:spice}
\end{figure}

Let us now focus on pretext tasks that are targeted to a specific downstream task. In the realm of music, two recent papers stand out, and greatly inspired our work. The first one is SPICE\footnote{SPICE stands for Self-Supervised Pitch Estimation} \citep{gfeller2020spice}. In their work, Gfeller et al. propose a pretext task that can be adapted to automatically estimate musical pitch. Two pitch-shifted pieces of the same audio are used as input to a same encoder-decoder network. Note that the CQT of each pitch-shifted track is used as input to the network. The encoder must then produce a single scalar for each CQT. This scalar is then used for two purposes. First, the relative difference between each scalar produced must be proportional to the initial pitch shift between each encoder input. Second, the scalar is used to reconstruct the un-shifted audio input. Both of these outputs are used in SPICE's loss function. The model proposed is then able to estimate pitch using a simple affine mapping, from relative to absolute pitch. Supervised learning is only used to calibrate this affine mapping. The results obtained on downstream pitch estimation tasks are superior to other fully-supervised methods, proving the efficacy of SPICE's pretext task. Figure \ref{fig:spice} \citep{gfeller2020spice} outlines the pretext task's full pipeline.

The second work that inspired us makes use of inverse audio synthesis to detect pitch. In their paper, Engel et al. \citep{engel2020self} make use of Differentiable Digital Signal Processing (DDSP) modules presented in \citep{engel2020ddsp}. An input log mel-spectrogram is reconstructed using a mixture of harmonic and sinusoidal synthesizers. By doing so, their network is able to disentangle a piece of music's pitch and timbre. The resulting pitch estimations outperform SPICE and other methods. To the best of our knowledge, these results are still state-of-the-art.



\subsection{Contrastive Learning}

Let us now introduce the key concepts behind contrastive learning. In some sense, contrastive learning is a more refined and modern version of triplet-based metric learning. We define anchors, positives, and negatives in the same way. Training batches are created using a sample pair (i.e. an anchor and its corresponding positive) and negatives that correspond to other samples' positives. These are generated randomly at each epoch. The standard contrastive loss function, defined by: 

\begin{equation}
    l_{a,p}=-log \left( \frac{exp(S(x_a, x_p)/T)}{\sum_{k=1, k \neq a}^{N} exp(S(x_a, x_k)/T)} \right)
    \label{eq:contra}
\end{equation}

is then computed across each batch during the training process. $S$ denotes a similarity function (most often Cosine Similarity) and $T$ a temperature parameter (usually between 0 and 1). We assume a batch size of $N$, where indices $a$ and $p$ are used for the anchor and positive.

\begin{figure}[H]
    \centering
    \includegraphics[width=1\textwidth]{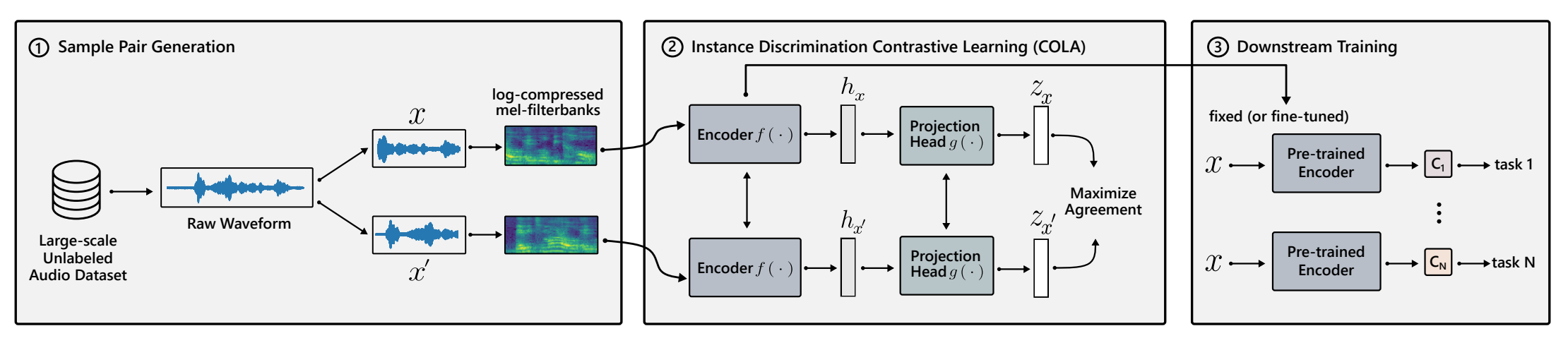}
    \caption{Overview of Contrastive Learning Applied to Audio \citep{saeed2021contrastive}}
    \label{fig:zeghi}
\end{figure}

In the field of audio, a few recent works inspired this thesis. First, Zeghidour et al. \citep{saeed2021contrastive} published a very simple framework for generating general-purpose audio representations. Anchors and positives are different sections of a same audio clip. Their log mel-spectrogram is then used as input to an EffcientNet model \citep{tan2019efficientnet} with two additional linear layers (used to project the output to a vector of size 512). The resulting outputs are compared using a contrastive loss with a bilinear similarity metric. The pre-trained EfficientNet model is then extracted and either frozen or fine-tuned for a set of downstream tasks that range from speaker identification to bird song detection. In almost all cases, the fine-tuned model vastly outperforms its fully-supervised counterpart. Figure \ref{fig:zeghi} outlines the simple framework used.

Another recent work in field was published by Wang et al. \citep{wang2021multi}. They use contrastive learning to match waveform audio representations to their spectrogram counterpart.

\begin{figure}[H]
    \centering
    \includegraphics[width=1\textwidth]{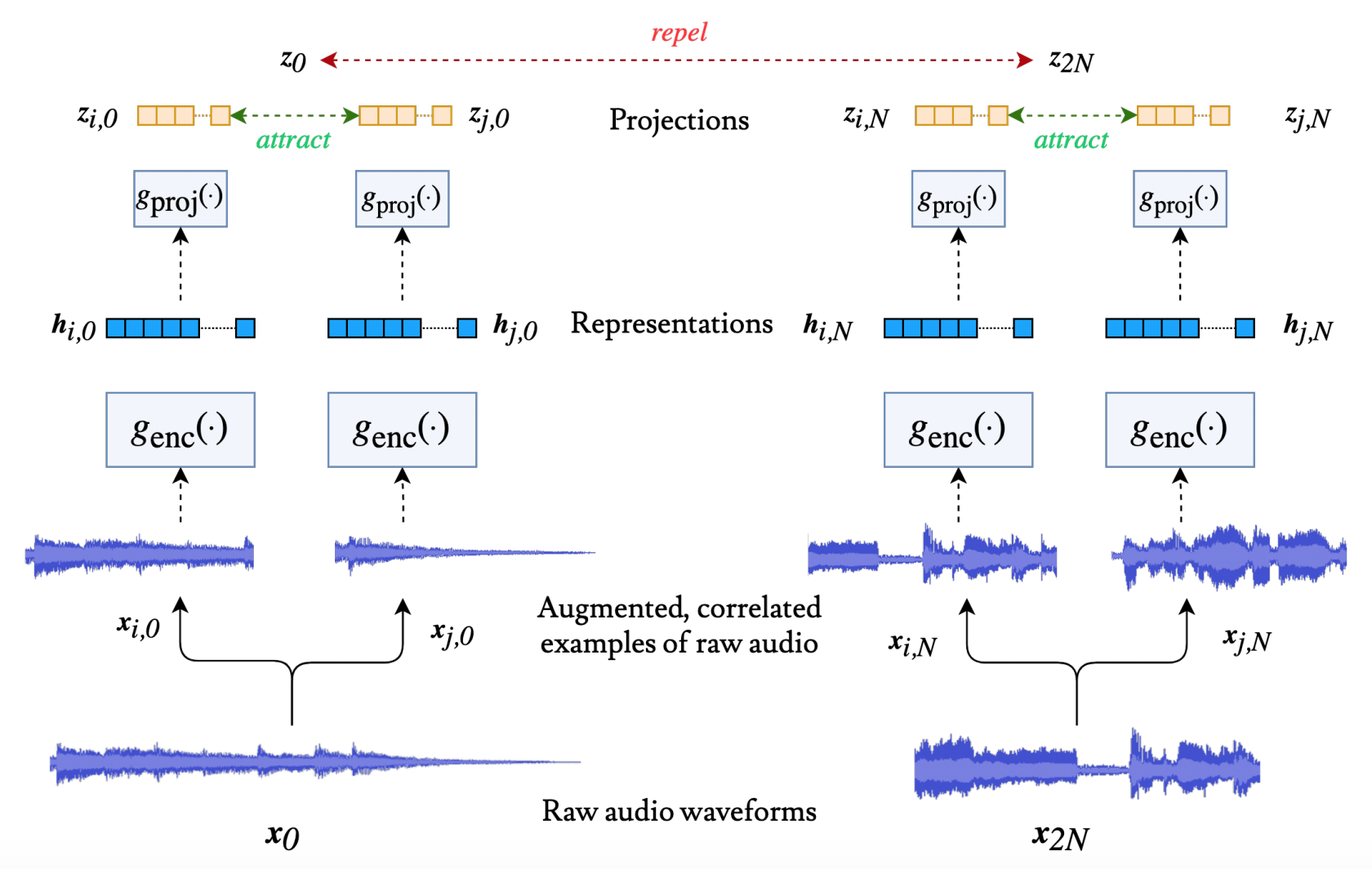}
    \caption{Overview of Contrastive Learning of Musical Representations \citep{spijkervet2021contrastive}}
    \label{fig:clmr}
\end{figure}

Finally, the most notable paper published in the field of musical contrastive learning is \textit{Contrastive Learning of Musical Representations} by Spijkervet et al. In their work, anchors and positives are waveform snippets from a same song. Positives are augmented using techniques such as polarity inversion and gain reduction. Their model is frozen and fine-tuned to a musical tagging downstream task using a fully-connected layer. The results they obtained are in-line with fully supervised methods at the time. More importantly, they achieve an extremely high performance using merely 1\% of the training data. Figure \ref{fig:clmr} outlines their pretext task pipeline.\footnote{Note that in both \citep{saeed2021contrastive} and \citep{spijkervet2021contrastive}, anchors are also compared to each other (i.e. are used as negatives for other anchors).}

\section{Audio Source Separation}

In audio, the field of blind source separation (BSS) deals with the task of recovering the source signals that compose a mixture. 
In this context, one does not know how many sources the mixture has. 
Many algorithms do however assume fixed sources to perform separation.\footnote{Spleeter \citep{spleeter2020}, for example, assumes 'drum,' 'bass,' 'vocal,' and 'other' sources in its '4stem' setting.} 
Historically, BSS tasks were solved using techniques from Auditory Scene Analysis (CASA) \citep{brown1994computational} or matrix decomposition methods. 
Most notably, independent component analysis (ICA) \citep{hyvarinen2000independent} was used to separate a mixture into statistically independent and non-Gaussian sources. 
\citep{ozerov2007adaptation} provides a more thorough overview of historical audio source separation methods.

In recent years, deep neural networks have taken over the field of source separation. The U-net architecture is extremely popular in the field of music, and has led to high-performing source separation algorithms \citep{jansson2017singing}. Other high-performing deep learning models in the field include \citep{stoller2018wave} and \citep{lluis2018end}. We encourage the reader to consult \citep{peeters2021deep} for a more thorough overview of deep learning networks applied to source separation. 

In this thesis, we use Spleeter\footnote{As a sidenote, Spleeter \citep{spleeter2020} was introduced by Deezer Research. Today, it is one of the most popular source separation algorithms in the field of music. It is notably used by large audio companies such as iZotope and Algoriddim. Note that the Python package is open-source.} \citep{spleeter2020} to separate our songs into multiple stems. Spleeter allows a user to split songs into two stems (vocal and other stems), four stems (vocal, bass, drum, and other stems), and five stems (vocal, bass, drum, piano, and other stems). We make use of the four-stem model to separate drum stems from the rest of our signals.

\section{Beat Tracking and Downbeat Estimation}

Let us finally introduce some of the important works in the fields of beat tracking and downbeat estimation. As mentioned previously, the beat of a song is often described as the rhythm a listener taps his foot to when listening to a piece of music. We refer to the downbeat estimation as the first beat of each bar. Before delving into methods, let us introduce some details about common datasets in the field. These are used in both our work and previous papers in the field.

\subsection{Datasets}

We use a total of four datasets to evaluate our beat tracking and downbeat estimation methodology. Table \ref{tab:data_sets} displays some of the important information about each of these (notably whether beat and downbeat annotations are available). 

\begin{table}[h]
\centering
\begin{tabular}{ |p{3.3cm}|p{1cm}|p{1.2cm}|p{1cm}|p{2.2cm}|  }
\hline
Dataset&\# files&length&Beats&Downbeats\\
\hline
Ballroom \citep{gouyon2006experimental} \citep{krebs2013rhythmic}&685&5h57m&yes&yes\\
Hainsworth \citep{hainsworth2004particle}&222&3h19m&yes&yes\\
GTZAN \citep{marchand2015swing} \citep{tzanetakis2002musical}&1000&8h20m&yes&yes\\
SMC \citep{holzapfel2012selective}&217&2h25m&yes&no\\
\hline
\end{tabular}
\caption{Common Datasets used for Beat Tracking and Downbeat Estimation}
\label{tab:data_sets}
\end{table}

On a separate note, the Ballroom dataset \citep{gouyon2006experimental} is comprised of dance music excerpts, such as tangos and waltzes. The Hainsworth \citep{hainsworth2004particle} dataset is comprised of a wider variety of genres, such as classical and electronic music. The GTZAN \citep{tzanetakis2002musical} dataset is comprised of 10 genres, spanning hiphop, jazz, and disco. Finally, the SMC \citep{holzapfel2012selective} dataset spans a wide range of genres that are similar to those of \citep{hainsworth2004particle}. One key difference with the other datasets is that each song was selected due the difficulty of estimating accurate beats. That is why most beat tracking systems perform much worst on this dataset than on others.

\subsection{Classic Methods}

Before deep learning, most methods in the field of beat tracking relied on a two-step process. The first was a front-end process that extracts onset locations (an onset describes the start of a musical event) from a time-frequency or subband analysis of a signal. A periodicity estimation algorithm would then find the rate at which these events occur. This is notably the case in \citep{miguel2004tempo}. By 2012, however, deep learning had already achieved state-of-the-art results in the field. We recommend the reader consult \citep{peeters2021deep} for a more thorough review of historical methods in beat tracking and downbeat estimation.

\subsection{Modern Methods}

When it comes to deep learning and beat estimation, a wide range of methods have been introduced lately. In these kinds of tasks, the neural network produces an activation function. This function is supposed to equal 1 when a beat occurs, and 0 otherwise. This function is then "picked" using a Dynamic Bayesian Network (DBN) \citep{krebs2015efficient} \citep{bock2014multi} \citep{krebs2013rhythmic}. These networks are probabilistic, and include Hidden Markov Models (HMM) and particle filtering models. They read the activation function and output its beat locations. We will come back to these later in this report. The first beat tracking architectures that were found to work were Recurrent Neural Networks (RNN) such as  Long Short Term Memory (LSTM) \citep{hochreiter1997long}. These types of networks are notably used in \citep{bock2014multi} and \citep{bock2016joint} to produce both beat and downbeat activation functions. More recently, temporal convolutional networks were found to perform just as well \citep{matthewdavies2019temporal}.

When it comes to training networks for beat tracking and/or downbeat estimation, we can distinguish two settings. The first is associated with training a model on one task only (i.e. we train one network on beat tracking or downbeat estimation only). In the second setting, both beat and downbeat locations are learned jointly during training, by two separate networks (the loss from each network is combined). This vastly improves results for downbeat estimation \citep{bock2016joint}. After all, beat tracking is an easier task, and a downbeat estimation network benefits from knowing where beats are located. Table \ref{tab:f1} summarizes the performance of some popular beat estimation methods. We report the papers' F1-measure\footnote{Note that, although the F1-measure is the most popular evaluation metric for beat tracking and downbeat estimation, a wide variety of other metrics also exist. We will briefly introduce them later in this report.} of correct versus incorrectly predicted beats. \textit{Correctness} is determined over a small window of 70 ms.

\begin{table}[h]
\begin{tabular}{ |p{3.3cm}|p{2.5cm}|p{1.5cm}|p{2.5cm}|  }
\hline
Dataset&Methodology&Beat F1&Downbeat F1\\
\hline
Ballroom \citep{gouyon2006experimental} \citep{krebs2013rhythmic}& TCN \citep{matthewdavies2019temporal} & 0.933 & NA \\
 & Joint RNN \citep{bock2016joint} & 0.938 & 0.863 \\
 & MM \citep{bock2014multi} & 0.910 & NA \\
 & FA-CNN \citep{durand2016feature} & NA & 0.778/0.797 \\
\hline
Hainsworth \citep{hainsworth2004particle}& TCN \citep{matthewdavies2019temporal} & 0.874 & NA \\
 & Joint RNN \citep{bock2016joint} & 0.867 & 0.684 \\
 & MM \citep{bock2014multi} & 0.843 & NA \\
 & FA-CNN \citep{durand2016feature} & NA & 0.657/0.664 \\
\hline
GTZAN \citep{marchand2015swing} \citep{tzanetakis2002musical}& TCN \citep{matthewdavies2019temporal} & 0.843 & NA \\
 & Joint RNN \citep{bock2016joint} & 0.856 & 0.640 \\
 & SPD \citep{davies2006spectral} & 0.806 & 0.462 \\
 & MM \citep{bock2014multi} & 0.864 & NA \\
 & FA-CNN \citep{durand2016feature} & NA & 0.860/0.879 \\
\hline
SMC \citep{holzapfel2012selective}& TCN \citep{matthewdavies2019temporal} & 0.543 & NA \\
 & Joint RNN \citep{bock2016joint} & 0.516 & NA \\
 & SPD \citep{davies2006spectral} & 0.337 & NA \\
 & MM \citep{bock2014multi} & 0.529 & NA \\
\hline
\end{tabular}
\caption{F-measures Obtained by Popular Beat Tracking and Downbeat Estimation Algorithms}
\label{tab:f1}
\end{table}

Most of the results obtained previously come from past \href{https://www.music-ir.org/mirex/wiki/2019:Audio_Beat_Tracking}{Mirex} challenges.\footnote{Some of the datasets that comprise the \href{https://www.music-ir.org/mirex/wiki/2019:Audio_Beat_Tracking}{Mirex} combined dataset are not available.} In these, the Ballroom \citep{gouyon2006experimental} \citep{krebs2013rhythmic}, Hainsworth \citep{hainsworth2004particle}, and SMC \citep{holzapfel2012selective} datasets are used for 8-fold Cross Validation (CV) whereas the GTZAN \citep{marchand2015swing} \citep{tzanetakis2002musical} dataset is used as a test set. We do not have access to the combined dataset, so evaluate our method using regular 8-fold CV on each dataset separately. This is a key difference that may explain the gap between some of the state-of-the-art results and ours.


\chapter{Methodology}\label{ch:mathtest} 
This chapter will cover the methods we used for our pretext and downstream task. 
The results we obtained will be presented in the next chapter.

\section{Audio Input Representation}

First, for both pretext and downstream tasks, all our audio input signals were resampled at a rate of 16000 Hz. 
In the case of the pretext task, this was done for all Spleeter-generated \citep{spleeter2020} stems. 
For the downstream tasks, this was done for each of our beat tracking data tracks. 

These signals were then transformed using Librosa's \cite{mcfee2015librosa} VQT. 
We used a hop length (number of audio samples between adjacent  Short-Time Frequency-Transform (STFT) columns) of 256. The minimum frequency used was 16.35 Hz (i.e. the frequency of the note C0). 
A total of 96 frequency bins were used for the resulting time-frequency representation. 
These correspond to a frequency range spanning eight octaves with a resolution of 12 notes per octave\footnote{All VQTs presented in this report follow the specifications described in this section.}. 
This VQT was inspired by the equal temperament tuning system. 
We then absolute valued and logged each bin.\footnote{Note that a small number $\epsilon = 10e-10$ was added to each bin before computing the $log$ operation.} 
The resulting matrix was used as input to all of our models.

\section{Model Design}

The model we designed was inspired by other beat tracking architectures \citep{bock2016joint} \citep{bock2014multi}.
Table \ref{tab:mod} summarizes the various layers that compose it. 
The table assumes an input shape of $96 \times 313$ (this corresponds to the VQT of five seconds of audio). 

The input log-VQT is first fed into a series of convolutional and max-pooling layers.
The max-pooling layers only diminish the frequency dimension.
The first max-pooling layer reduces the dimension from 96 to 32, the second from 32 to 8, and the third from 8 to 1. 
In some sense, this is akin to reducing the frequency dimension to one value per octave, and max-pooling the resulting values.
The time dimension is not reduced, however. 
This is due to the fact that our input time dimension already has a resolution of 62.6 bins per second, or approximately 16 ms per bin. 
For tasks such as beat tracking, this time resolution is standard. 
In \citep{matthewdavies2019temporal}, for example, the authors use a time resolution of 10 ms per bin to achieve state-of-the-art results.

\begin{table}[h]
\begin{tabular}{ |p{2cm}|p{3.3cm}|p{1.3cm}|p{1.3cm}|p{1.5cm}|  }
\hline
Layer & Output Dimension \linebreak (\# Channels $\times$ Freq. \linebreak Bins $\times$ Time Dim.) & Kernel Size & Stride & Padding\\ 
\hline
Input& 1 $\times$ 96 $\times$ 313 & & & \\
\hline
Conv2d& 64 $\times$ 96 $\times$ 313 & 3 $\times$ 11 & 1 $\times$ 1 & 1 $\times$ 5 \\
MaxPool2d& 64 $\times$ 32 $\times$ 313 & 3 $\times$ 1 & 3 $\times$ 1 & 0 $\times$ 0 \\
ReLU& & & & \\
DropOut& & & & \\
\hline
Conv2d& 128 $\times$ 32 $\times$ 313 & 5 $\times$ 15 & 1 $\times$ 1 & 2 $\times$ 7 \\
MaxPool2d& 128 $\times$ 8 $\times$ 313 & 4 $\times$ 1 & 4 $\times$ 1 & 0 $\times$ 0 \\
ReLU& & & & \\
DropOut& & & & \\
\hline
Conv2d& 256 $\times$ 8 $\times$ 313 & 3 $\times$ 21 & 1 $\times$ 1 & 1 $\times$ 10 \\
MaxPool2d& 256 $\times$ 1 $\times$ 313 & 8 $\times$ 1 & 8 $\times$ 1 & 0 $\times$ 0 \\
ReLU& & & & \\
DropOut& & & & \\
\hline
Conv2d& 128 $\times$ 1 $\times$ 313 & 1 $\times$ 25 & 1 $\times$ 1 & 0 $\times$ 12 \\
ReLU& & & & \\
DropOut& & & & \\
\hline
GRU& 256 $\times$ 313 & & & \\
DropOut& & & & \\
\hline
Conv1d& 1 $\times$ 313 & 1 & 1 & 0 \\
Sigmoid& & & & \\
\hline
\end{tabular}
\caption{Model Architecture}
\label{tab:mod}
\end{table}

The convolutional layers' kernel sizes are widened on the time dimension as the network deepens. 
The number of channels is also increased up to 256. 
Once the frequency dimension size is equal to 1, the sequence is fed into a stacked GRU \citep{cho2014learning} layer (i.e. two consecutive GRUs; the latter reads the output of the former). 
Each GRU reads sequences in a bidirectional fashion. 
The stacked GRU output is then fed into a convolutional layer with kernel size $1 \times 1$.
This layer reduces the number of channels to 1.
A sigmoid activation layer is then used to squash all values between 0 and 1.
The resulting sequence corresponds to our model's output.

Do note that all convolutional and GRU layers are followed by a combination of ReLU activations and Dropout. 
For our pretext task, we used a Dropout probability value of 0.1 (values superior to 0.3 made it impossible for our model to train). 
This value was set to 0.5 for our downstream tasks. 
This was done so that our model would train correctly in the former case, and to combat overfitting in the latter case.
Furthermore, we originally chose a Dropout value of 0.5 due to its optimality for a wide range of tasks \citep{srivastava2014dropout}.

\section{Pretext Task}

Let us now introduce the pipeline we created for our contrastive learning experiment.

\subsection{Data Processing}

The first step of our experiment was to create a very large dataset that contained snippets of drum and Rest-of-Signal (ROS) snippets. 
Note that we define a ROS stem as a track without its drums (for simplicity purposes).

In order to do so, we first loaded each track in the FMA large \citep{defferrard2016fma} dataset (106,574 tracks of 30s) using a re-sampling rate of 44100 Hz (Spleeter assumes an input sample rate of 44100 Hz).
We then used Spleeter \citep{spleeter2020} to separate each track into its drum and ROS stems. 
The latter was created by mixing the \textit{bass}, \textit{other}, and \textit{vocals} stems extracted by Spleeter's \textit{4stems} model. 
Once this step was done, we were presented with a set of problem: not all tracks contain drums of course, and matching a ROS stem to an empty signal would likely make our pretext task fail (especially if the training set contains multiple tracks without drums, which is likely the case). 
We also ran into numerous cases where our drum stem contained all the audio, while our ROS stem was empty. 
This often occurred when a track was comprised of a bassline and its accompanying drums. 
When both were synchronized, the source separation model struggled to disentangle them.

In order to solve this problem, we computed the Root-Mean-Square (RMS), defined for a signal $x$ of length $n$ as:

\begin{equation}
    x_{RMS} = \sqrt{\frac{1}{n}(x_{1}^2 + x_{2}^2 +...+x_{n}^2)}
\end{equation}

on both the ROS and drum stems. 
Note that the RMS was computed over frames of length $n=2048$ using a hop length of 512. 
This allowed us to verify whether drums occurred throughout the extracted stem, and not just in short sections at the beginning or end.

Assuming $N$ RMS values were extracted for each stem, we then verified that the following inequality was verified, for $k \in [1, N]$:

\begin{equation}
   \frac{ROS_{RMS}(k)}{2} < DRUM_{RMS}(k) < ROS_{RMS}(k) * 4
   \label{eq:rms}
\end{equation}

\begin{figure}[h]
\centering
\subfloat[][Spleeter Drum Stem]{\includegraphics[width=0.5\textwidth]{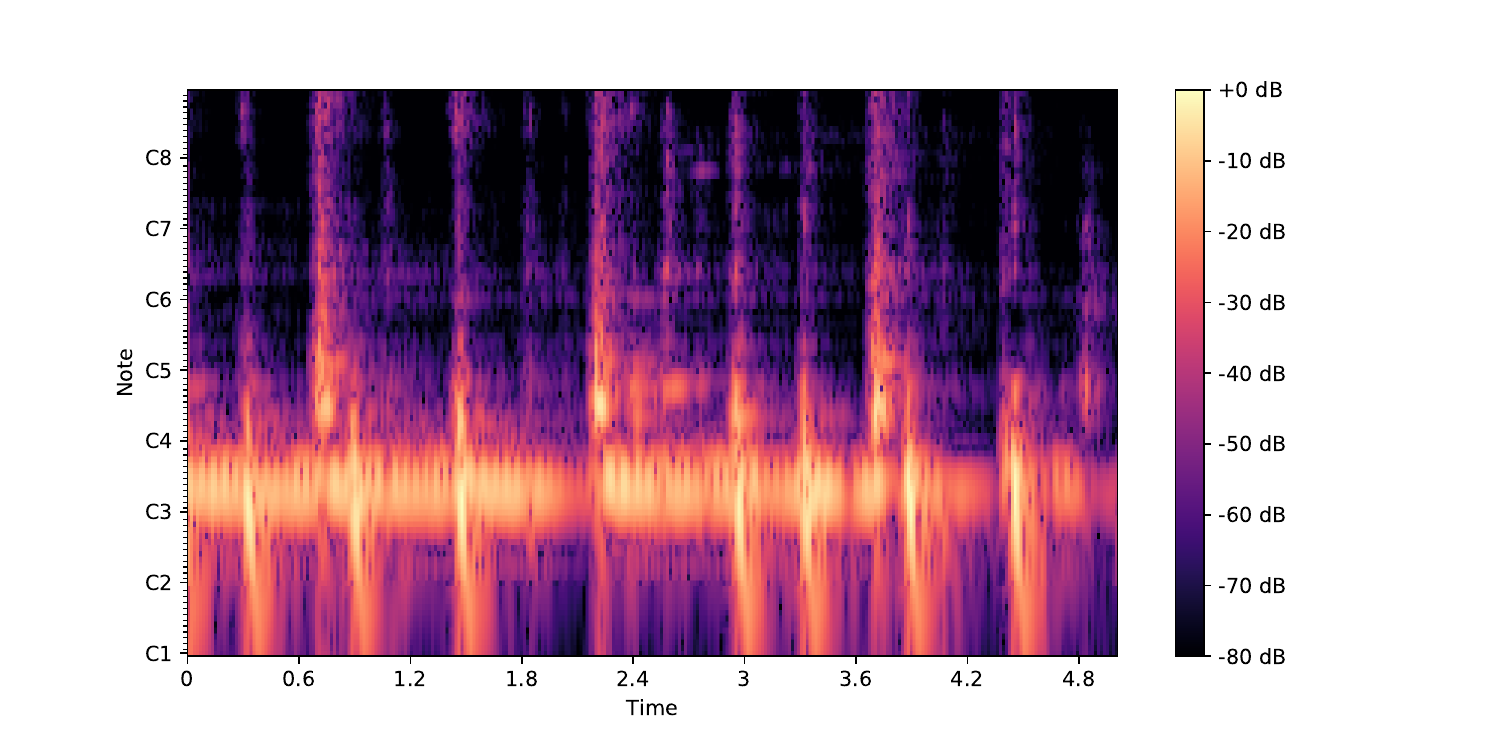}}
\subfloat[][Ground Truth Drum Stem]{\includegraphics[width=0.5\textwidth]{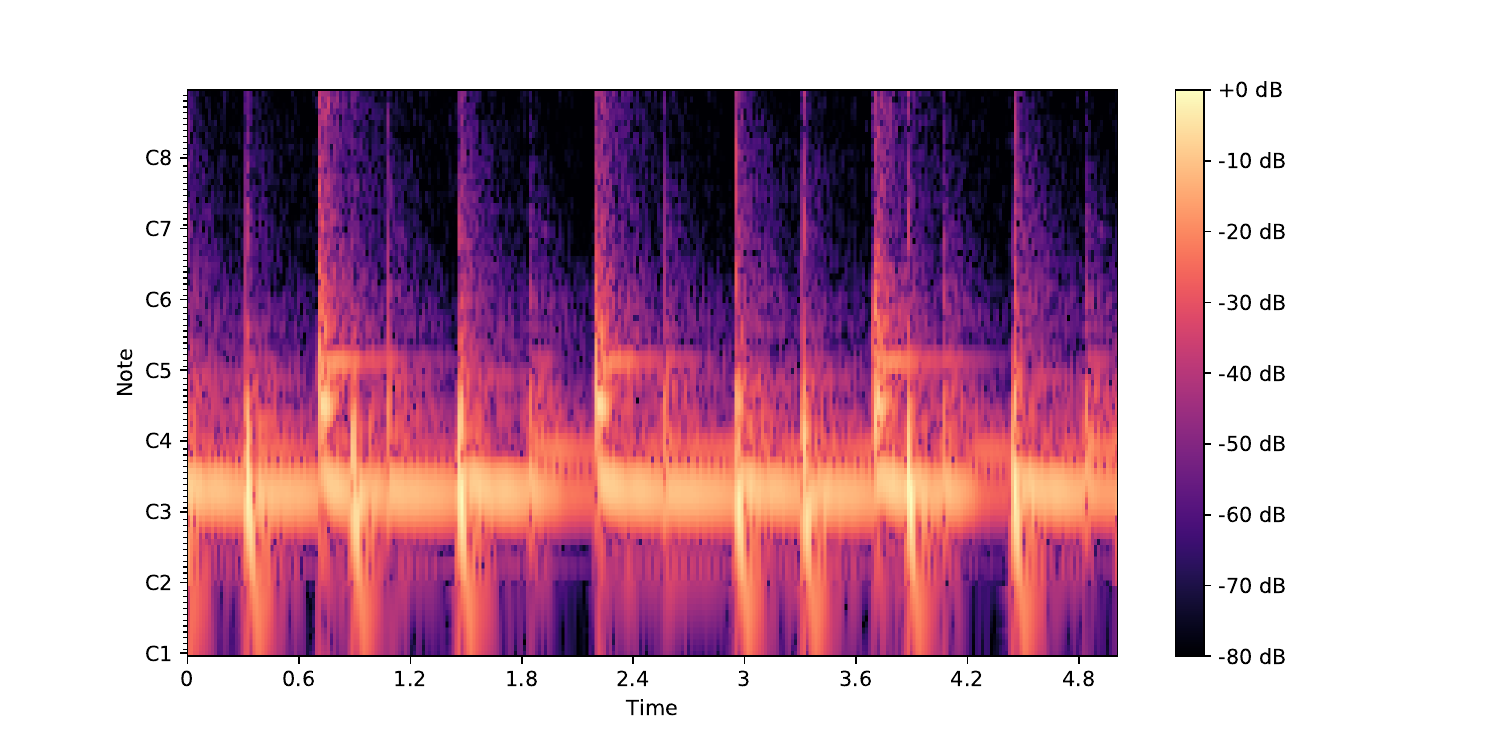}}\\
\subfloat[][Spleeter Mixture of Other Stems]{\includegraphics[width=0.5\textwidth]{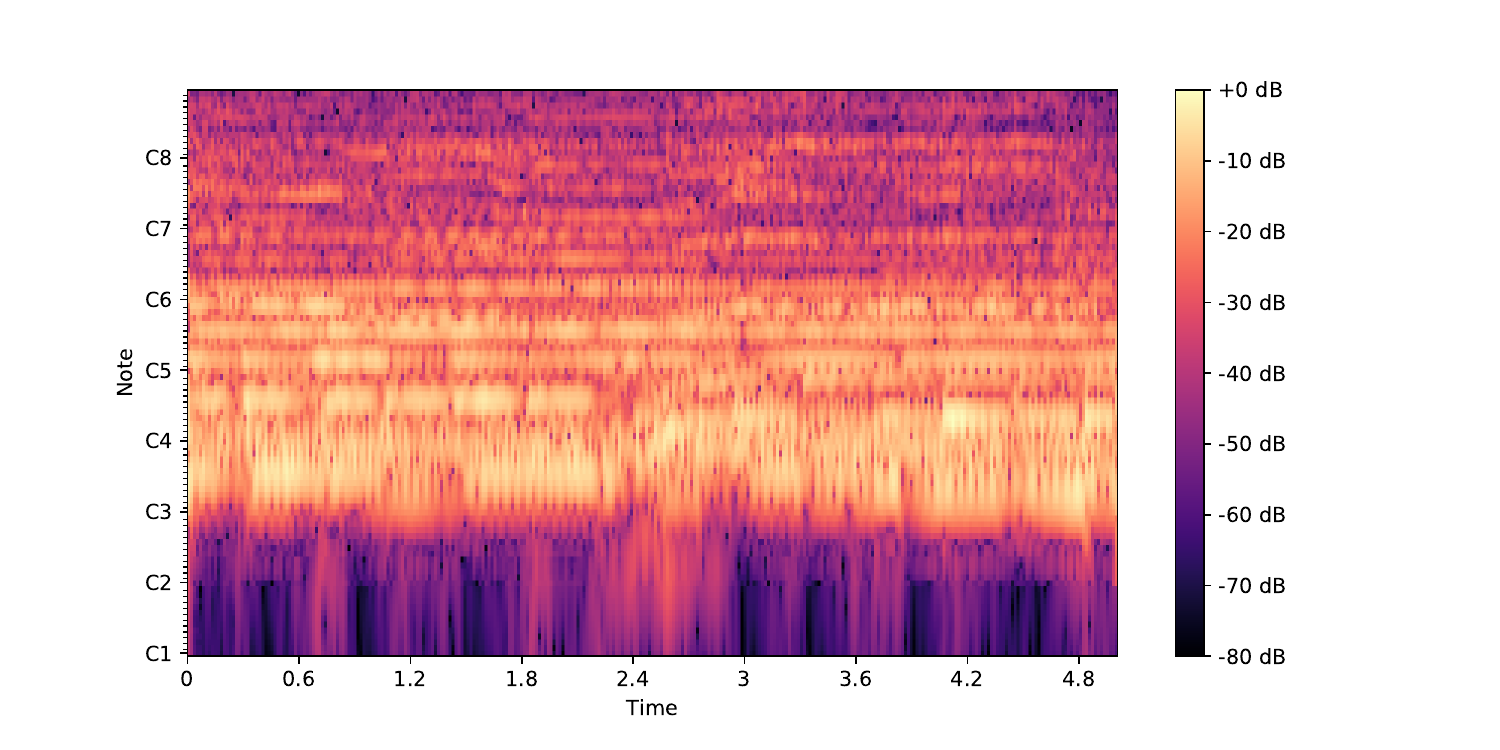}}
\subfloat[][Ground Truth Mixture of Other Stems]{\includegraphics[width=0.5\textwidth]{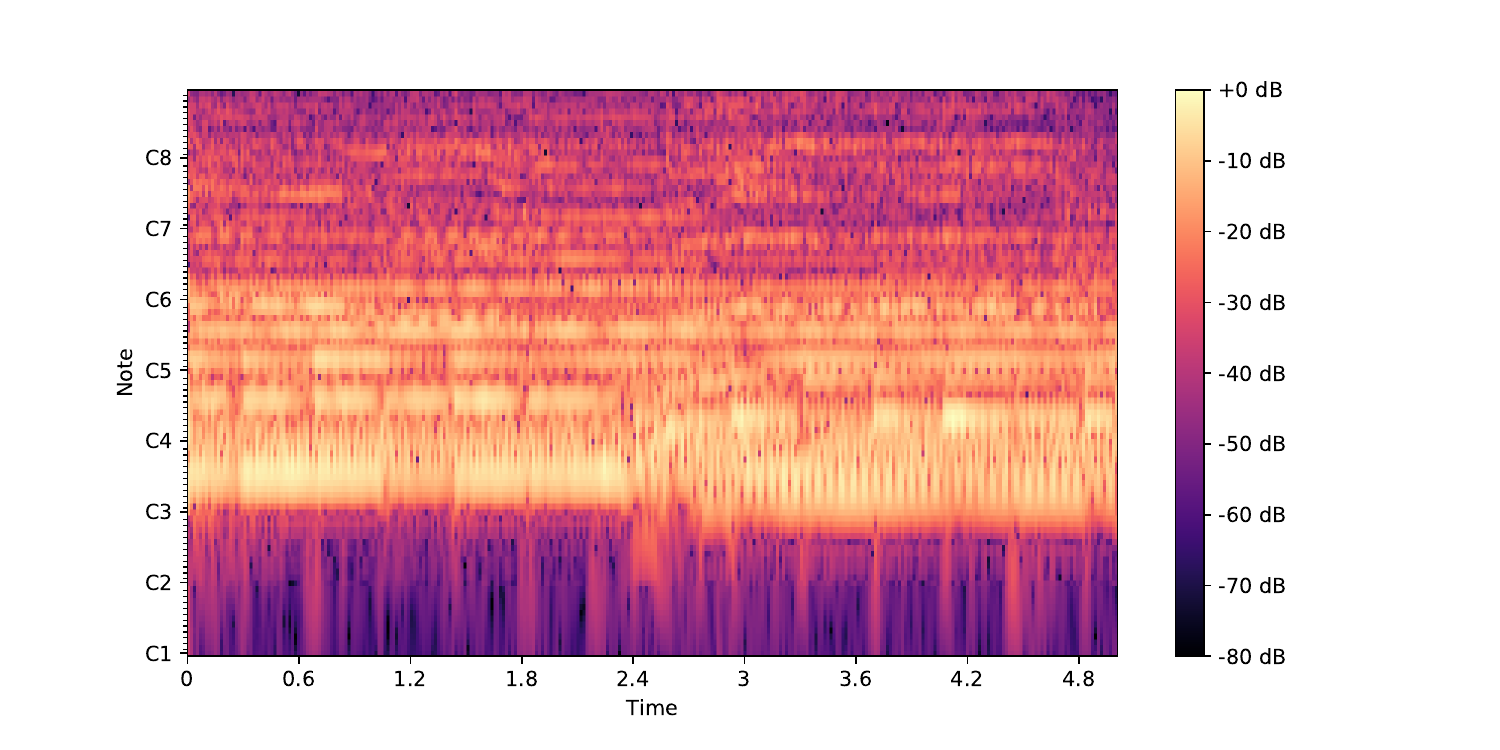}}\\
\subfloat[][Full Signal]{\includegraphics[width=0.5\textwidth]{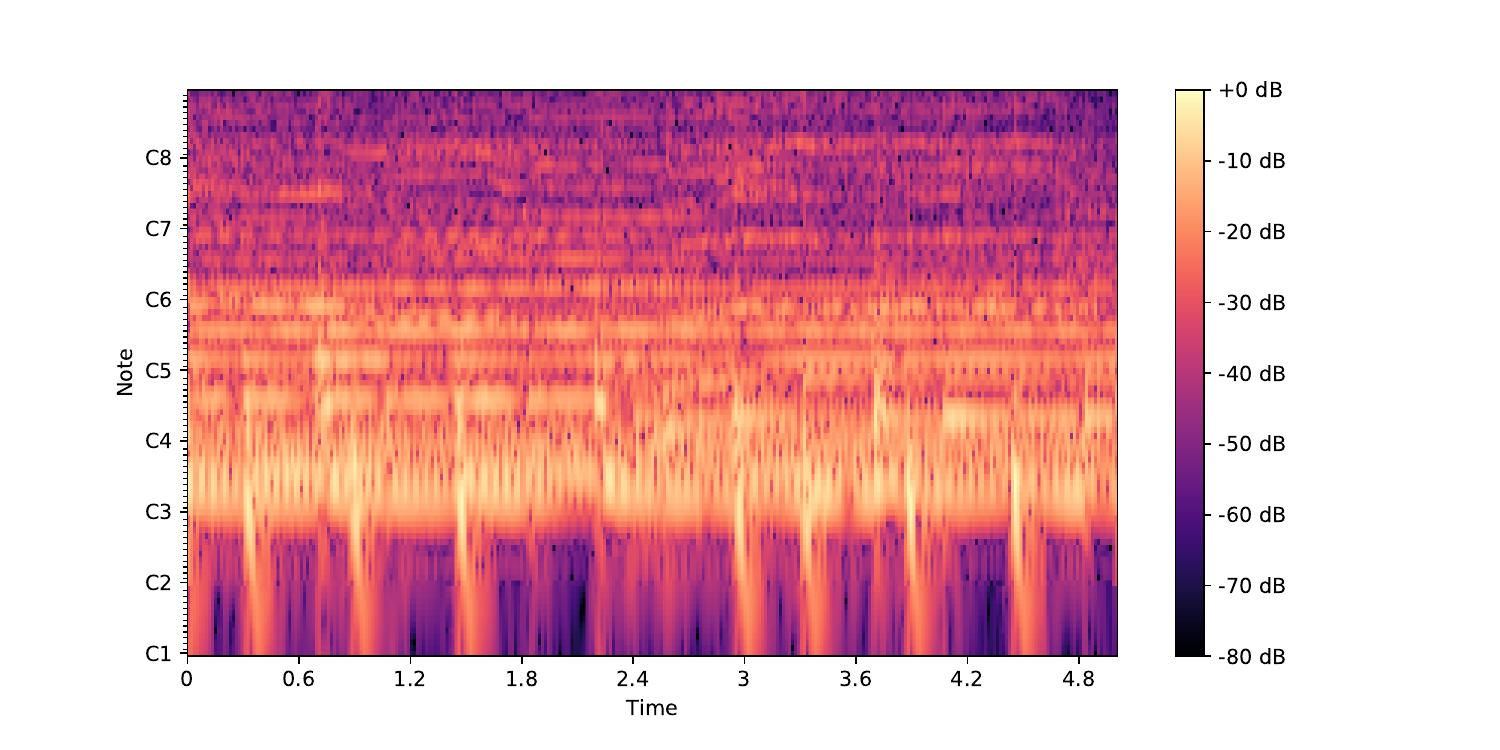}}
\caption[VQT Representations of Our Model Inputs]{VQT representations of our model inputs. We also provide VQT illustrations of the ground truth drum and ROS stems. 
As one can notice, the Spleeter-generated stems are similar to the ground truths. 
The drum stem does however seem to contain artefacts, especially at the higher-frequency level. 
We will come back to the role these artefacts may have played in our pretext task in subsequent chapters.}
\label{fig:stems}
\end{figure}

If the above was satisfied for over 30\% of the $N$ values obtained for each stem, 5 seconds of the song's audio verifying \ref{eq:rms} were extracted and used in our pretext training set\footnote{One final note: we chose to save our pretext data VQTs in memory before training our model. 
Our data processing can of course be done on-the-fly during training, but loading, re-sampling, and transforming audio to its VQT representation is a very expensive process time-wise.}. 
Note that the process above was found to work very well in-practice, and allowed us to obtain training set stems that were balanced (i.e. had nicely separated drum and ROS signals). 
\footnote{This may have contributed to a song selection bias.
We will come back to this idea in Chapter 6.}
A total of 35200 5-second stem pairs were generated from the FMA \citep{defferrard2016fma} dataset in this way. 
This corresponds to almost 49 hours of audio. 
The VQT of each stem was then computed. 
Figure \ref{fig:stems}\footnote{We apologize for the size of the $x$ and $y$ axes for some figures in this chapter and the next. Unfortunately, many of these were generated during our experiments, and would take an experiment re-run to recreate.} illustrates our pretext task inputs.

\subsection{Batch Creation}

Once all 35200 sample pairs were created, we created our batches by first randomly selecting an anchor (ROS stem VQT) and its corresponding positive (drum stem VQT). We then filled the rest of the batch using other randomly selected positives. For a batch size of 64, this corresponds to 62 other drum stems. 

We split our data into a training set of size 28800 and a validation set of size 6400\footnote{We randomly selected anchors at each validation epoch. 
We acknowledge this may be bad practice, as the validation set batches were not constant throughout training. 
Regardless, our model exhibited the desired behaviour during the learning process.}. 
Hence, during each epoch, 450 anchors were used for training and 100 for validation. 
The contrastive loss defined in the previous section, \ref{eq:contra}, was then computed over each batch and used to train the model.

Note that we used Cosine Similarity, defined by:

\begin{equation}
    cosim(a, b) = \frac{a \cdot b}{||a|| ||b||} = \frac{\sum_{i=1}^{n}a_i b_i}{\sqrt{\sum_{i=1}^{n}a_i^2}\sqrt{\sum_{i=1}^{n}b_i^2}}
\end{equation}

for two vectors $a$ and $b$, as a similarity metric. Similar vectors have a Cosine Similarity close to 1, whereas dissimilar vectors have a Cosine Similarity close to -1.

\subsection{Experimental Setup}

Our model and loss function were computed using PyTorch \citep{NEURIPS2019_9015}. 
We used the Adam optimizer \citep{kingma2014adam} throughout training with an initial learning rate of $5e-6$\footnote{Higher learning rates did not enable our model to train. 
Even worse, they often led to an exploding gradients problem.}. 
This learning rate was divided by 2 every 200 epochs. We stopped training the model after 425 epochs\footnote{We initially planned to train our model for 600 epochs. Due to memory issues, training was stopped at 425 epochs. The model's training plots exhibited the desired behaviour so we opted to stop any further training.}. 
The model that performed the best on the validation set (i.e. that had the lowest average loss across each validation batch) was saved and re-used for our downstream tasks.

\section{Downstream Tasks}

\subsection{Data Processing}

As outlined earlier, we used four datasets for our beat tracking task and three for our downbeat estimation task. The Ballroom \citep{gouyon2006experimental} \citep{krebs2013rhythmic}\footnote{Note that the duplicates in the Ballroom \citep{gouyon2006experimental} \citep{krebs2013rhythmic} dataset identified in \url{https://highnoongmt.wordpress.com/2014/01/23/ballroom_dataset/} were removed.}, Hainsworth \citep{hainsworth2004particle}, and GTZAN \citep{marchand2015swing} \citep{tzanetakis2002musical} datasets were used for both downstream tasks whereas the SMC \citep{holzapfel2012selective} dataset was used for beat tracking only.

\begin{figure}[h]
    \centering
    \includegraphics[width=1.\textwidth]{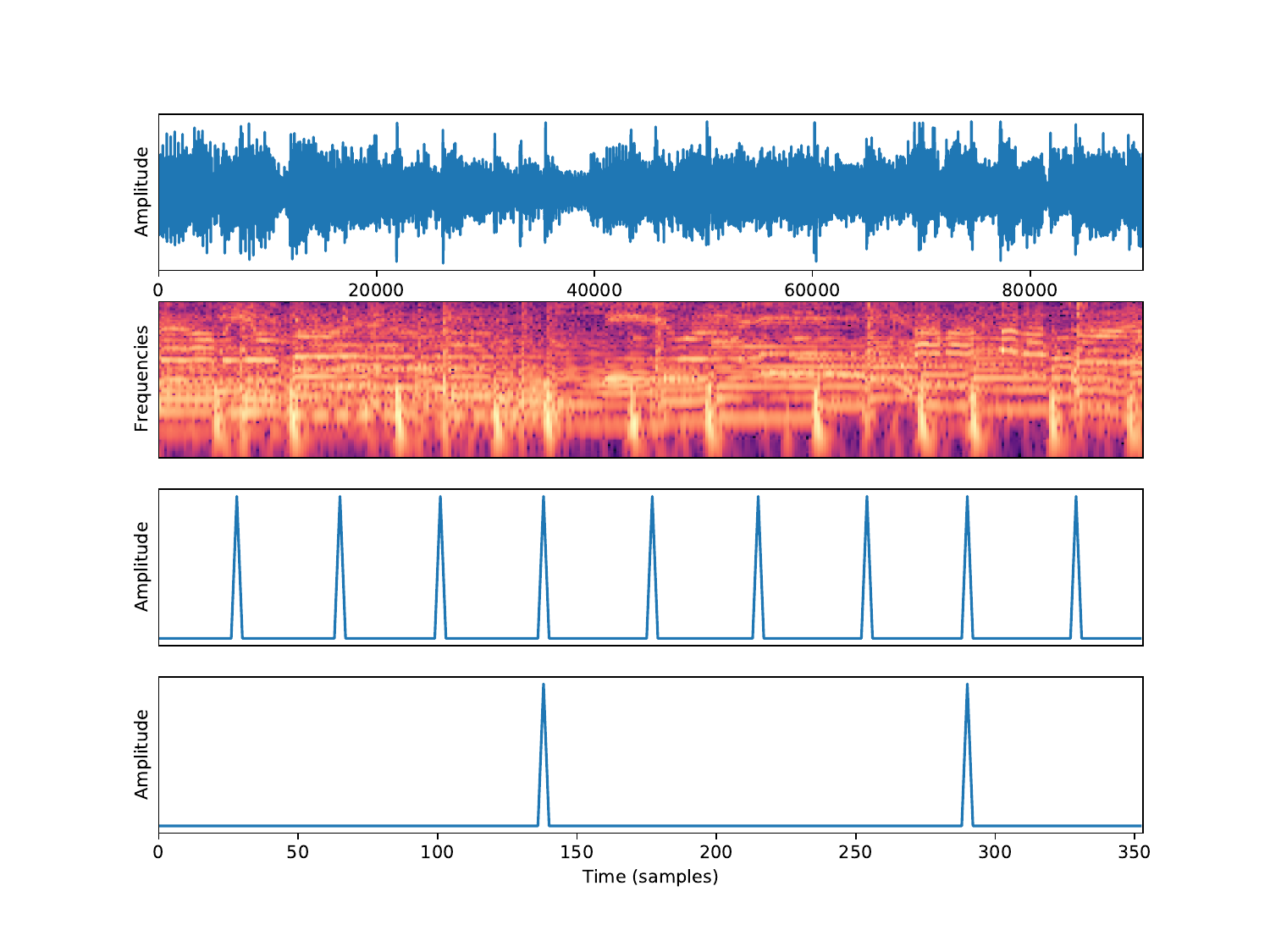}
    \caption[Beat Tracking Data Processing Pipeline]
	{Beat Tracking Data Processing Pipeline. The first figure represents the input song's waveform representation, the second its VQT representation, the third its beat tracking activation function, and the fourth its downbeat estimation activation function.}
    \label{fig:beat}
\end{figure}

For each song in the datasets, we computed the VQT using the same resolution as the pretext task. 
It was used as input to our model. 
The target output of our model was an activation function with the same time resolution as our input VQT. 
We used the annotations provided by each dataset to determine the locations of each beat (i.e. each value of 1) on the activation function. 
Note that the time steps preceding and following each beat were annotated with a value of 0.5 to aid the model in its beat identification task.

Figure \ref{fig:beat} outlines the data processing pipeline we used, with examples of both the input and target outputs of our model.

\subsection{Experiments}

Let us now introduce the experiments we conducted. 
For each of these, the model was initialized either randomly or with the pretext task's pre-trained weights. 

Note that for every experiment, the Binary Cross-Entropy (BCE) loss between the target and model output, defined by:

\begin{equation}
    L = -\frac{1}{N}\sum_{i=1}^{N} y_i \cdot log(x_i) + (1 - y_i) \cdot log(1 - x_i)
\end{equation}

for targets $y$ and predictions $x$ of length $N$, was computed.



\begin{table}[h]
\begin{tabular}{ |p{3.3cm}|p{3.3cm}|p{3.9cm}|  }
\hline
dataset&Vanilla Learning&Pre-trained Learning\\
\hline
Ballroom \citep{gouyon2006experimental} \citep{krebs2013rhythmic}& & \\
- Beat & $5e-5$ & $5e-5$ \\
- Downbeat & $1e-4$ & $5e-5$ \\
- Joint & $1e-4$ & $1e-4$ \\
\hline
Hainsworth \citep{hainsworth2004particle}& & \\
- Beat & $1e-4$ & $5e-5$ \\
- Downbeat & $1e-4$ & $5e-5$ \\
- Joint & $1e-4$ & $5e-5$ \\
\hline
GTZAN \citep{marchand2015swing} \citep{tzanetakis2002musical}& & \\
- Beat & $5e-5$ & $5e-5$ \\
- Downbeat & $1e-4$ & $5e-5$ \\
- Joint & $2e-4$ & $1e-4$ \\
\hline
SMC \citep{holzapfel2012selective}& & \\
- Beat & $1e-4$ & $5e-5$ \\
\hline
\end{tabular}
\caption{Experimental Setup Learning Rates}
\label{tab:lrs}
\end{table}

Table \ref{tab:lrs} outlines the learning rates used for the experiments described in the subsequent sub-sections. 
All our models were trained for a maximum of 50 epochs. These were then evaluated using the F1-score, AMLc, AMLt, CMLc, and CMLt evaluation metrics. The CMLc and CMlt metrics evaluate how continuously correct a beat tracking estimation is (use of the maximum length of correct predictions). The AMLt and AMLc metrics are similar but allow offbeat variations of an annotated beat sequence to be matched with detected beats. One can read more about each metric in \citep{davies2009evaluation}.

\newpage

\paragraph{Pure Beat Tracking and Downbeat Estimation}\mbox{}\\

The first, and simplest, experiments we conducted were aimed at training our model for either one of our downstream tasks. 
For both beat tracking and downbeat estimation dataset, we used 8-fold CV to evaluate our model. 
Each fold was used once as a test set. The rest of the data samples were used for training or validation.\footnote{For each test fold, the validation set was comprised of randomly selected data samples from the other seven folds. 
Note that both the test and validation sets were of the same size.} 
For each experiment, we used the Adam \citep{kingma2014adam} optimizer with learning rates described in Table \ref{tab:lrs} and a batch size of 1.\footnote{We used a batch size of 1 because most tracks did not have the same input size.} 
The model that achieved the highest mean F1-score on the validation set over 50 epochs was selected for evaluating the test set.
This was done for each of the 8 test sets that compose a dataset.
We evaluated every single dataset separately.

The pre-trained DBN in \citep{krebs2015efficient} was used to "pick" our activation function for both beat tracking and downbeat estimation. 
Since it is tailored to beat tracking tasks, we used a Beats-per-Minute (BPM) range of $[55, 215]$ for the former task, and $[10, 75]$ in the latter case. 
These ranges were found to be optimal for each task and every dataset. 
Note that all DBNs were provided by the madmom \citep{bock2016madmom} Python library. 
All the evaluation metrics were computed using the $mir\_eval$ \citep{raffel2014mir_eval} library and its default settings. 

Finally, the learning rates used were generally in the same range. 
For larger datasets, we used the same learning rate for both vanilla and fine-tuning training. 
Models were found to train correctly in both instances. 
For smaller datasets, we often divided the learning rate by two for fine-tuning, as the model would quickly overfit. 
This was done because pre-training was found to greatly accelerate learning. 
The next chapter will cover this in-depth.

\paragraph{Joint Estimation}\mbox{}\\

In this experimental setting, our network's goal is to learn beat and downbeat annotations in parallel. 
Beat tracking is usually an easier task, and can guide a downbeat estimation network.
This form of learning usually greatly enhances a model's downbeat estimation capabilities.
We defined two networks with the same architecture described previously. 
One focused on beat tracking whilst the other focused on downbeat estimation. 
BCE loss was computed on each of the two networks' outputs. 
The sum of both losses was then backpropagated to the networks.

We used the DBN defined in \citep{bock2016joint}\footnote{Note that we limited our DBN's beats-per-bar setting to 3 and 4 (i.e. the DBN only models bars with 3 or 4 beats) for our joint estimation task.} to process both activations simultaneously. 
This allowed our beat and downbeat outputs to be synchronized in time. 
At each epoch, we summed the mean F1-scores obtained by each model on the beat tracking and downbeat estimation validation set. 
The models that achieved the highest sum were selected for testing. 
The rest of the experimental setup was exactly like in the previous section.


\paragraph{Impact of Training Set Size on Learning Performance}\mbox{}\\

In order to verify the impact of our pre-training on downstream performance, we studied the amount of training examples needed to achieve a decent performance on the joint estimation task. 
We isolated an eighth of each one of our datasets as test sets. 
Another eighth was used for our validation sets. 
From there, a random subset of the remaining data were used as our training set. 
This subset corresponded to 1\%, 2\%, 5\%, 10\%, 20\%, 50\%, or 75\% of the remaining training set. 
Table \ref{tab:percs} outlines the size of the train sets for each dataset.

We ran this random selection and training process 10 times for each percentage of the train set used. 
The test set results were then averaged to determine whether pre-trained models needed less data to achieve higher performance.

\paragraph{Cross-dataset Generalization}\mbox{}\\

The final experiment we conducted was centred around determining how our models generalized from one dataset to another.\footnote{8-fold CV was also applied. Each fold was used as a validation set.} We trained our models on one of the GTZAN \citep{tzanetakis2002musical} \citep{marchand2015swing} or Ballroom \citep{gouyon2006experimental} \citep{krebs2013rhythmic} datasets, and tested them on the Hainsworth dataset \citep{hainsworth2004particle}. This was only done in a joint estimation setting.


\chapter{Results}\label{ch:mathtest} 
Let us now explore the results we obtained for each one of our experiments. We will start by analyzing our pretext task, through the lens of both the training behaviour our model displayed and the resulting onset function. We will then analyze the effects of pre-training on our beat tracking and downbeat estimation tasks. For each downstream task, both vanilla and pre-trained model performances will be reported. The goal of these experiments was to:

\begin{enumerate}
    \item determine whether our pretext task could help our model's performance on beat tracking and downbeat estimation tasks.
    \item determine whether our pretext task could help train our model using less labelled data.
    \item determine whether our pretext task could help our downstream models generalize their performance to new datasets.
\end{enumerate}

\section{Pretext Task}

\subsection{Training Behaviour}

When training our neural network on our computer-generated data set, one can notice that the loss decreases slowly but surely.\footnote{Do note that our model was tailored to perform well on the downstream tasks too. Popular deep neural networks, such as Residual Networks \citep{he2016deep}, performed much better on our pretext task, but were not suited to our downstream tasks.} The mean batch loss does not however decrease very drastically (it only decreases from an initial value of 4.1 to approximately 3.5). 

The evolution of the cosine similarity metric is however quite interesting. We obtain the behaviour we initially desired: anchors and positives whose Cosine Similarity increases towards 1, and anchors and negatives whose Cosine Similarity gradually decreases towards 0\footnote{Since our model's output vectors are positive, the minimum possible distance between our vectors is 0.} (the mean similarity between anchors and negatives plateaus at around 0.2). These values are averages of all the 550 batches present in our training and validation sets.\footnote{Since our dataset contains more than one sample pair, achieving a perfect anchor/ positive similarity of 1 and  a perfect anchor/ negative similarity of 0 is almost impossible.} More importantly, both sets exhibit similar behaviour, suggesting that our model is indeed capable of matching the correct drum and ROS stems throughout each batch. 

\begin{figure}[H]
\subfloat[][Mean Batch Loss Evolution. The blue line represents the mean train set batch loss for each epoch, whereas the orange one represents the mean validation set batch loss.]{\includegraphics[width=1.\textwidth]{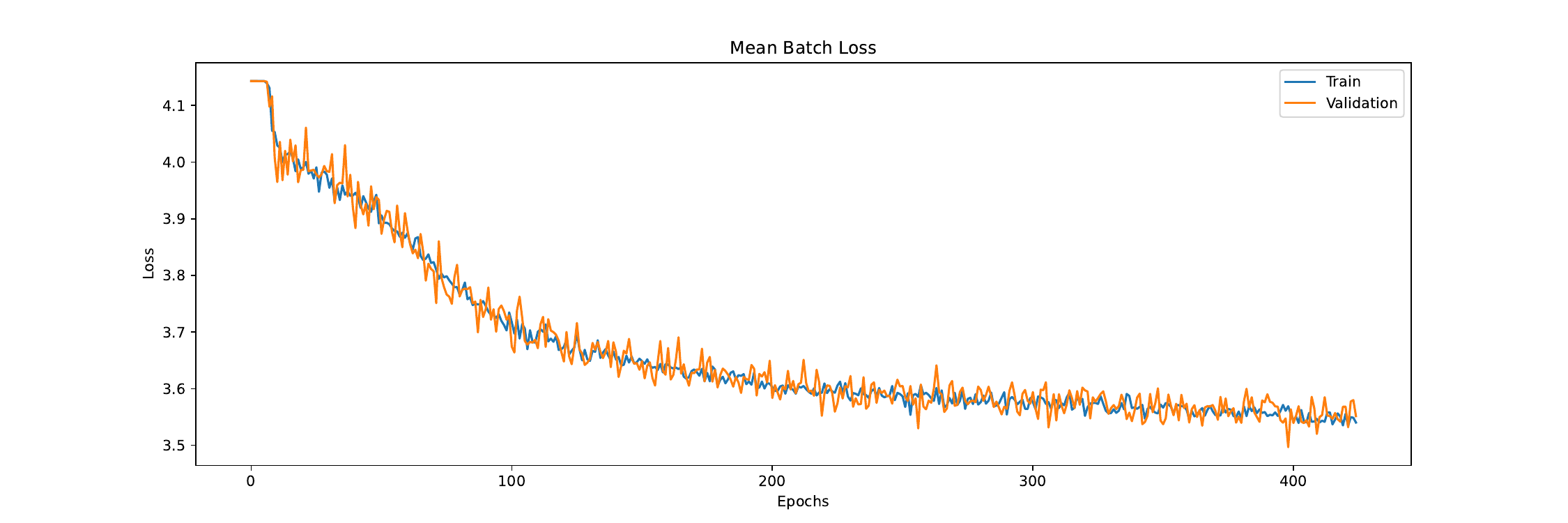}}\\
\subfloat[][Mean Train Set Cosine Similarity. The blue line represents the Cosine Similarity between anchors and positives, whereas the orange one represents the Cosine Similarity between anchors and negatives.]{\includegraphics[width=1.\textwidth]{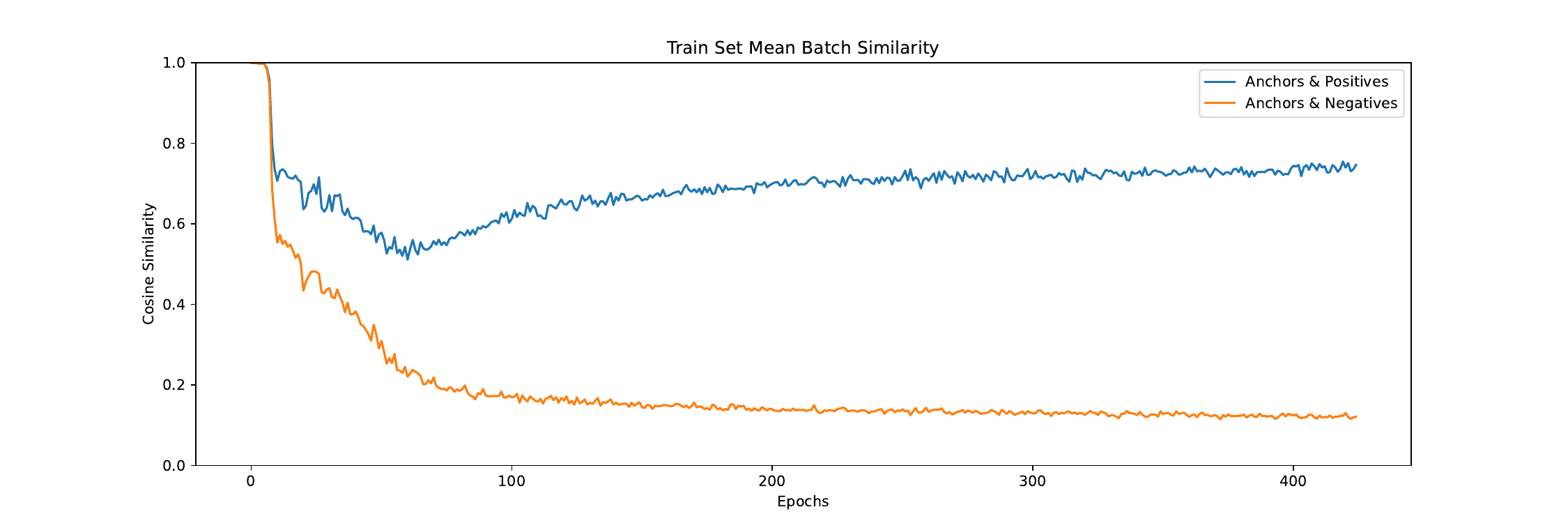}}\\
\subfloat[][Mean Validation Set Cosine Similarity. The blue line represents the Cosine Similarity between anchors and positives, whereas the orange one represents the Cosine Similarity between anchors and negatives.]{\includegraphics[width=1.\textwidth]{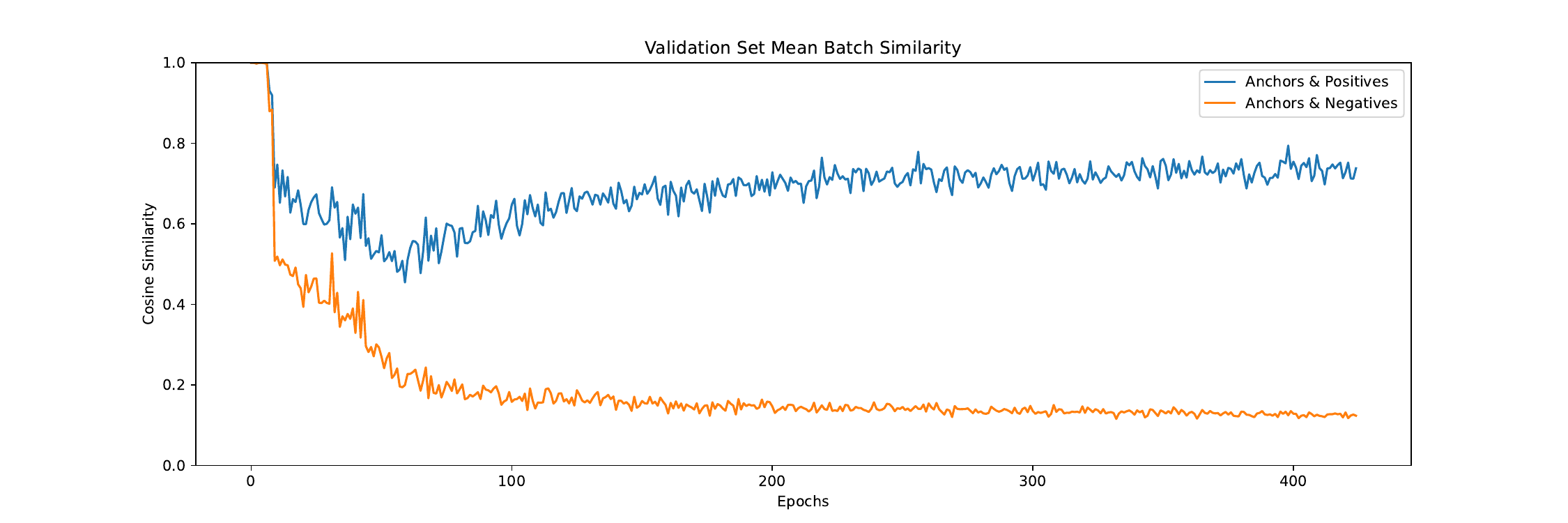}}
\caption{Pretext Task Training Behaviour}
\label{fig:convergence}
\end{figure}

\subsection{Onset Function}

When analyzing the vectorial representation learned by our model, one can immediately notice that it greatly resembles an onset function. For the most part, the vector's values are close to 0. They do however "spike" during certain musical events. We evaluated our model on the Mus dB \citep{musdb18} data set to judge its performance. This data set is comprised of 150 songs, and each of their stems. We extracted 10-seconds worth of drum and ROS audio for each stem in order to gauge our pre-trained network's performance on longer audio segments.

\begin{figure}[h]
\centering
\subfloat[][Superimposed Anchor and Positive Onset Functions]{\includegraphics[width=1.\textwidth]{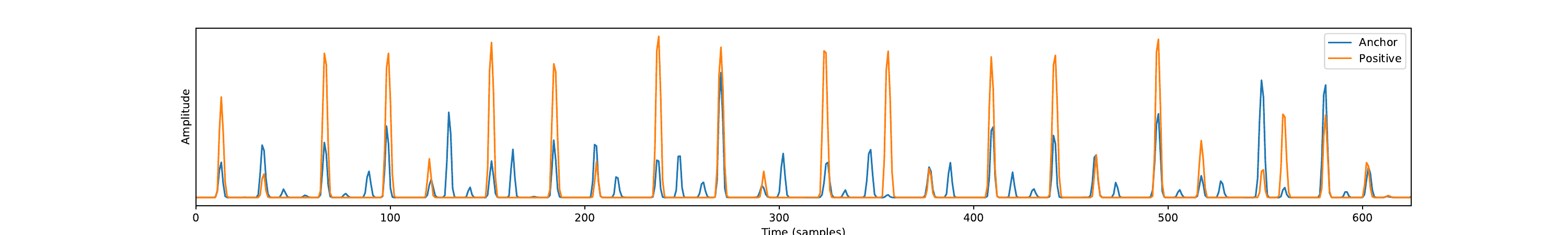}}\\
\subfloat[][Superimposed Positive Signal and Onset Function]{\includegraphics[width=1.\textwidth]{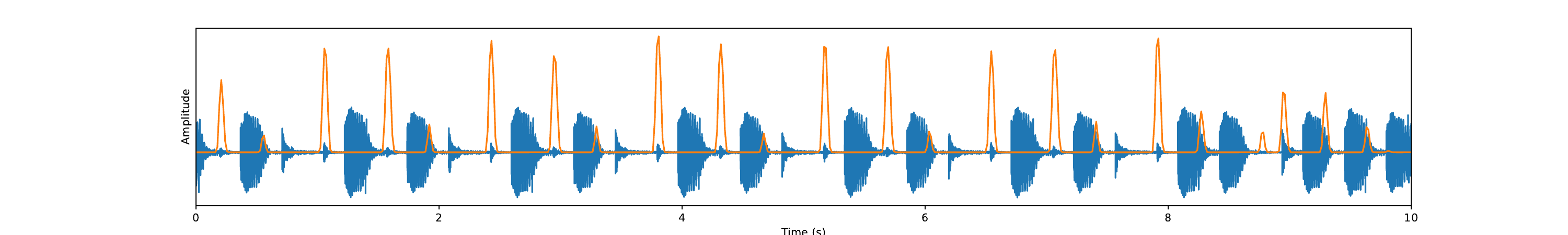}}\\
\subfloat[][Superimposed Anchor Signal and Onset Function]{\includegraphics[width=1.\textwidth]{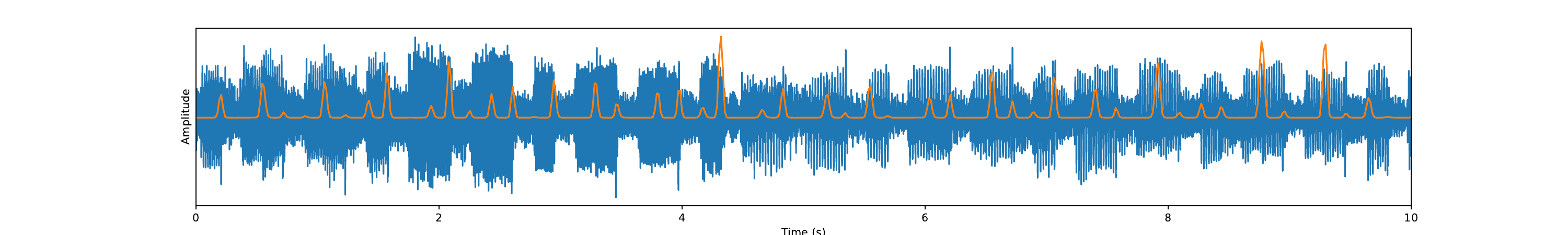}}
\caption[Successful Stem Match]{Successful Stem Match. The Cosine Similarity between the anchor and positive is 0.663.}
\label{fig:success}
\end{figure}

\begin{figure}[H]
\centering
\subfloat[][Superimposed Anchor and Positive Onset Functions]{\includegraphics[width=1.\textwidth]{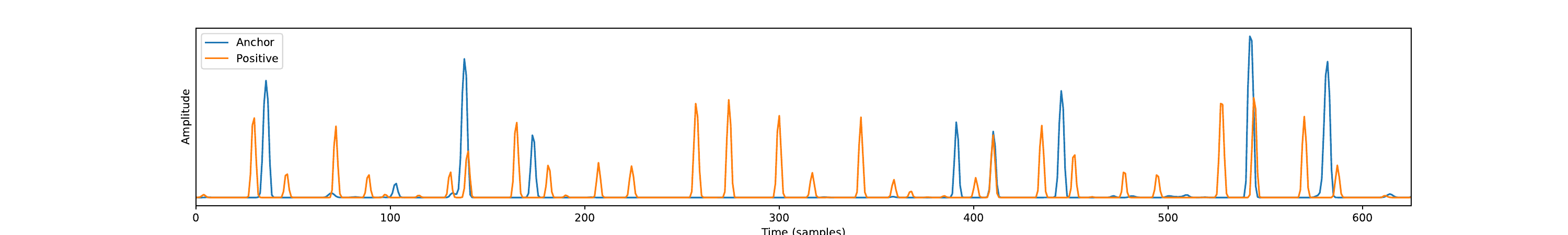}}\\
\subfloat[][Superimposed Positive Signal and Onset Function]{\includegraphics[width=1.\textwidth]{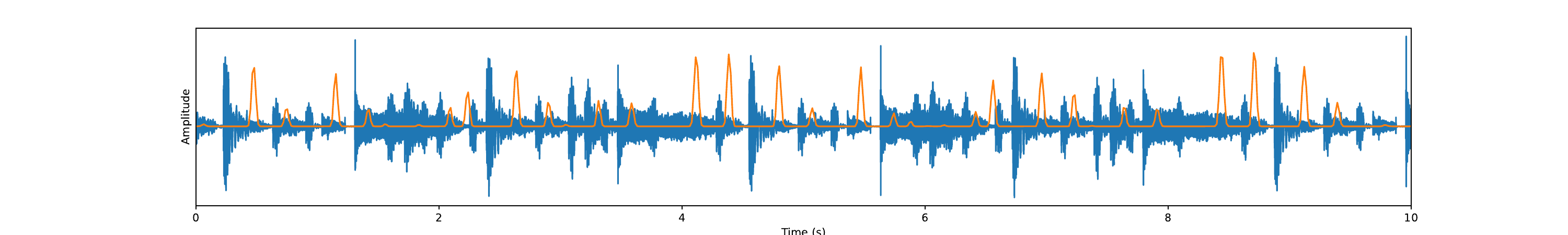}}\\
\subfloat[][Superimposed Anchor Signal and Onset Function]{\includegraphics[width=1.\textwidth]{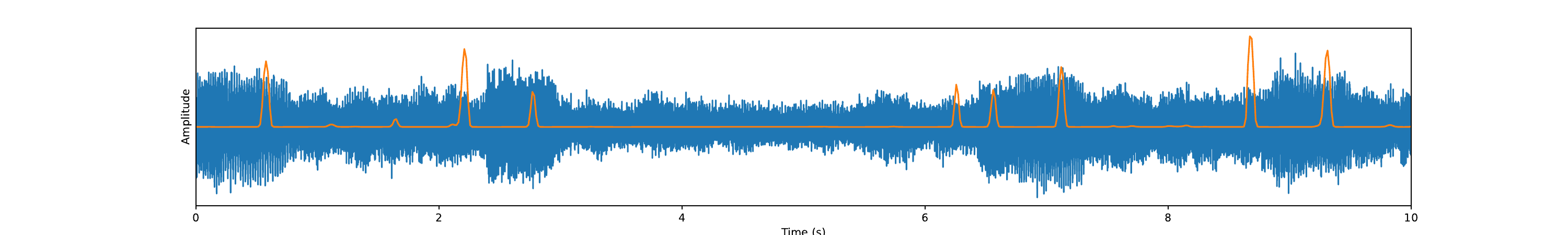}}
\caption[Failed Stem Match]{Failed Stem Match. The Cosine Similarity between the anchor and positive is 0.101.}
\label{fig:fail}
\end{figure}

When computing the drum and ROS stem representations of these clean stems, we first noticed that the performance was somewhat lower. Successful matches only had a Cosine Similarity in the range of $[0.4, 0.6]$ (compared to 0.7+ using Spleeter-generated \citep{spleeter2020} stems), and a number of sample pairs had a similarity closer to 0.1. This suggests that Spleeter's artifacts \citep{spleeter2020} may have aided the model's matching of anchors and positives.\footnote{Visually, it seems like Spleeter \citep{spleeter2020} mostly created spectral "holes" in the anchor's VQT throughout the source separation process. These holes were usually drum locations and helped our model match ROS and drum stems. Note that this remains a hypothesis.}

Moreover, the representations learned by the model were found to resemble an onset function. This is due to the fact that output vectors are quite parsimonious. The observed "spikes" seem to correlate nicely with rhythm, however. This is notably the case in Figure \ref{fig:success} (b), where the peaks seem to be aligned with the kick sounds (they are just shifted in time). Further work needs to be done in order to determine whether these onset functons can be used in a standalone\footnote{By standalone, we refer to the idea that the onset functions would be used as the sole input to a MIR algorithm.} fashion for tasks such as beat tracking or tempo estimation. Figure \ref{fig:fail} illustrates a failed stem match. In this case, the onset functions are much less interpretable.

\section{Downstream Tasks}

Let us now study our downstream task performance. The following sub-section contains tables outlining our various results. These were compared to the state-of-the-art methods in the field. For each table, we report the mean and standard deviation test set performance (usually F1-measure). The standard deviation values were not provided by any other papers.

\paragraph{Pure Beat Tracking and Downbeat Estimation}\mbox{}\\

When looking at our results (Table \ref{tab:res_beat}), a few elements stand out. 
First, our network's performance on beat tracking tasks is quite good compared to other state-of-the-art methods, using both random and pre-trained initializations. 
This is especially the case for larger datasets such as GTZAN \citep{marchand2015swing} \citep{tzanetakis2002musical} and Ballroom \citep{gouyon2006experimental} \citep{krebs2013rhythmic}. 
This is not the case on the Hainsworth dataset \citep{hainsworth2004particle} however. The Hainsworth data set \citep{hainsworth2004particle} is quite small, and benefits greatly from being trained alongside other data sets. Bock et al. \citep{matthewdavies2019temporal} \citep{bock2016joint} do so in their works, whereas we train a different model on each dataset.\footnote{And for each fold.} This most likely explains the performance gap observed.

Second, pre-training does not help with performance. In fact, in most cases, the network's performance slightly worsens with pre-training. Note that we tried using smaller learning rates, larger drop-out, and frozen layers to no avail. We believe this is due to the fact that pre-training led our network to overfit more easily. Figures \ref{fig:ball}-\ref{fig:hain}-\ref{fig:gt} in the appendix illustrate this idea. When the network is pre-trained, we observe that the validation set's F1-score is higher during the first few epochs. This most likely led our network to cater to the training set too fast, and by extension generalize less well to unseen data.

Finally, the model's performance on the pure downbeat estimation task is extremely poor compared to the state-of-the-art today (it is still quite good compared to previous methods). Downbeat estimation is a very complex task which benefits greatly from knowing a song's beat annotations. Our results in the next subsection demonstrate this idea. Tables \ref{tab:res_beat} and \ref{tab:res_down} outline the mean test set performance using 8-fold CV for each dataset.

\begin{table}[H]
\centering
\begin{tabular}{ |p{3cm}|p{1.1cm}|p{1.1cm}|p{1.1cm}|p{1.1cm}|p{1.1cm}|}
\hline
Dataset&\textit{F1}&\textit{CMLc}&\textit{CMLt}&\textit{AMLc}&\textit{AMLt}\\
\hline
\textit{Ballroom \citep{gouyon2006experimental} \citep{krebs2013rhythmic}}& & & & &\\
\textit{- Vanilla}& \textbf{0.933 (0.011)} & \textbf{0.865 (0.019)} & \textbf{0.884 (0.020)} & 0.908 (0.008) & \textbf{0.929 (0.008)}\\
\textit{- Pre-trained}& 0.920 (0.011) & 0.854 (0.022) & 0.872 (0.021) & 0.896 (0.015) & 0.916 (0.014)\\
\textit{- TCN \citep{matthewdavies2019temporal}} & \textbf{0.933} & 0.864 & 0.881 & \textbf{0.909} & \textbf{0.929}\\
\hline
\textit{Hainsworth \citep{hainsworth2004particle}}& & & & &\\
\textit{- Vanilla}& 0.753 (0.029) & 0.556 (0.057) & 0.627 (0.051) & 0.752 (0.078) & 0.849 (0.063)\\
\textit{- Pre-trained}& 0.757 (0.041) & 0.533 (0.083) & 0.600 (0.088) & 0.748 (0.040) & 0.845 (0.030)\\
\textit{- TCN \citep{matthewdavies2019temporal}} & \textbf{0.874} & \textbf{0.755} & \textbf{0.795} & \textbf{0.882} & \textbf{0.930}\\
\hline
\textit{GTZAN \citep{marchand2015swing} \citep{tzanetakis2002musical}}& & & & &\\
\textit{- Vanilla}& \textbf{0.862 (0.022)} & \textbf{0.748 (0.045)} & \textbf{0.771 (0.039)} & 0.866 (0.032) & 0.899 (0.024)\\
\textit{- Pre-trained}& 0.859 (0.019) & 0.737 (0.035) & 0.760 (0.032) & 0.876 (0.028) & 0.906 (0.027)\\
\textit{- TCN \citep{matthewdavies2019temporal}} & 0.843 & 0.695 & 0.715 & \textbf{0.889} & \textbf{0.914}\\
\hline
\textit{SMC \citep{holzapfel2012selective}}& & & & &\\
\textit{- Vanilla}& 0.528 (0.027) & \textbf{0.346 (0.062)} & \textbf{0.452 (0.073)} & \textbf{0.473 (0.018)} & 0.628 (0.030)\\
\textit{- Pre-trained}& 0.526 (0.057) & 0.337 (0.084) & 0.451 (0.092) & 0.447 (0.081) & 0.610 (0.080)\\
\textit{- TCN \citep{matthewdavies2019temporal}} & \textbf{0.543} & 0.315 & 0.432 & 0.462 & \textbf{0.632} \\
\hline
\end{tabular}
\caption[Pure Beat Tracking Results]{Pure Beat Tracking Results. We compare our results with those in \citep{matthewdavies2019temporal}. Like us, they make use of a CNN and DBN to obtain their results in a supervised fashion. The performance on larger datasets, such as GTZAN \citep{marchand2015swing} \citep{tzanetakis2002musical} and Ballroom, match or outperform \citep{matthewdavies2019temporal}. This is not the case for the Hainsworth \citep{hainsworth2004particle} and SMC \citep{holzapfel2012selective} datasets. Also, the vanilla model often outperforms the pre-trained model. 8-fold CV standard deviation is reported between parentheses.}
\label{tab:res_beat}
\end{table}

\begin{table}[H]
\centering
\begin{tabular}{ |p{3cm}|p{1.1cm}|p{1.1cm}|p{1.1cm}|p{1.1cm}|p{1.1cm}|}
\hline
Dataset&\textit{F1}&\textit{CMLc}&\textit{CMLt}&\textit{AMLc}&\textit{AMLt}\\
\hline
\textit{Ballroom \citep{gouyon2006experimental} \citep{krebs2013rhythmic}}& & & & &\\
\textit{- Vanilla}& 0.570 (0.024) & 0.090 (0.036) & 0.090 (0.036) & 0.597 (0.035) & 0.603 (0.033)\\
\textit{- Pre-trained}& 0.557 (0.013) & 0.090 (0.020) & 0.090 (0.020) & 0.583 (0.045) & 0.588 (0.043)\\
\textit{- Joint RNN \citep{bock2016joint}} & \textbf{0.863} & NA & NA & NA & NA\\
\hline
\textit{Hainsworth \citep{hainsworth2004particle}}& & & & &\\
\textit{- Vanilla}& 0.481 (0.079) & 0.294 (0.097) & 0.302 (0.101) & 0.643 (0.108) & 0.663 (0.101)\\
\textit{- Pre-trained}& 0.492 (0.063) & 0.276 (0.069) & 0.284 (0.070) & 0.685 (0.075) & 0.701 (0.076)\\
\textit{- Joint RNN \citep{bock2016joint}} & \textbf{0.684} & NA & NA & NA & NA\\
\hline
\textit{GTZAN \citep{marchand2015swing} \citep{tzanetakis2002musical}}& & & & &\\
\textit{- Vanilla}& 0.460 (0.008) & 0.021 (0.007) & 0.022 (0.007) & 0.433 (0.029) & 0.442 (0.026)\\
\textit{- Pre-trained}& 0.445 (0.017) & 0.015 (0.008) & 0.016 (0.009) & 0.420 (0.033) & 0.429 (0.033)\\
\textit{- Joint RNN \citep{bock2016joint}} & \textbf{0.640} & NA & NA & NA & NA\\
\hline
\end{tabular}
\caption[Pure Downbeat Estimation Results]{Pure Downbeat Estimation Results. We compare these results with \citep{bock2016joint}. Although this paper also makes use of a CNN and DBN, the RNN is trained in parallel with a beat tracking network. It therefore significantly outperforms our network, which was trained purely for the downbeat estimation task. Again, we observe that pre-training our network did not lead to a significant performance gain on any of our datasets.}
\label{tab:res_down}
\end{table}

\paragraph{Joint Estimation}\mbox{}\\

In our joint estimation setup, the effects of pre-training were more noticeable. This is most likely due to the fact that two networks were initialized with pre-trained weights. Whilst vanilla learning struggled with training both networks simultaneously (especially during the first 5-10 epochs), our pretext task training allowed the model to achieve better results more quickly. Overall, pre-training our network led to better test set performance for both downbeat estimation and beat tracking. Do however note that the obtained results are not yet up-to-par with state-of-the-art joint estimation methods. Tables \ref{tab:joint_beat} and \ref{tab:joint_down} contain the experiment's results and commentary.

One should also notice how much better the downbeat estimation results are. In the pure downbeat estimation task, our model never achieved a mean F1-score above 0.570 on the Ballroom \citep{gouyon2006experimental} \citep{krebs2013rhythmic} dataset, 0.492 on the Hainsworth dataset \citep{hainsworth2004particle}, and 0.460 on the GTZAN dataset \citep{tzanetakis2002musical} \citep{marchand2015swing}. In the joint setup, we obtain maximum scores of 0.822, 0.517, and 0.613. The model in both experiments does not change. However, the joint training method allows it to learn much better.

\begin{table}[H]
\centering
\begin{tabular}{ |p{3cm}|p{1.1cm}|p{1.1cm}|p{1.1cm}|p{1.1cm}|p{1.1cm}|}
\hline
Dataset&\textit{F1}&\textit{CMLc}&\textit{CMLt}&\textit{AMLc}&\textit{AMLt}\\
\hline
\textit{Ballroom \citep{gouyon2006experimental} \citep{krebs2013rhythmic}}& & & & &\\
\textit{- Vanilla}& 0.885 (0.049) & 0.744 (0.097) & 0.769 (0.102) & 0.861 (0.058) & 0.889 (0.059)\\
\textit{- Pre-trained}& 0.909 (0.023) & \textbf{0.795 (0.059)} & \textbf{0.826 (0.064)} & \textbf{0.872 (0.019)} & \textbf{0.905 (0.021)}\\
\textit{- Joint RNN \citep{bock2016joint}} & \textbf{0.938} & NA & NA & NA & NA\\
\hline
\textit{Hainsworth \citep{hainsworth2004particle}}& & & & &\\
\textit{- Vanilla}& 0.763 (0.062) & \textbf{0.579 (0.074)} & \textbf{0.657 (0.077)} & \textbf{0.717 (0.087)} & \textbf{0.818 (0.078)}\\
\textit{- Pre-trained}& 0.750 (0.052) & 0.541 (0.073) & 0.626 (0.071) & 0.692 (0.056) & 0.798 (0.050)\\
\textit{- Joint RNN \citep{bock2016joint}} & \textbf{0.867} & NA & NA & NA & NA\\
\hline
\textit{GTZAN \citep{marchand2015swing} \citep{tzanetakis2002musical}}& & & & &\\
\textit{- Vanilla}& 0.829 (0.017) & 0.661 (0.044) & 0.690 (0.038) & 0.833 (0.026) & 0.870 (0.020)\\
\textit{- Pre-trained}& 0.831 (0.022) & \textbf{0.675 (0.037)} & \textbf{0.702 (0.031)} & \textbf{0.842 (0.030)} & \textbf{0.877 (0.022)}\\
\textit{- Joint RNN \citep{bock2016joint}} & \textbf{0.856} & NA & NA & NA & NA\\
\hline
\end{tabular}
\caption[Joint Estimation Beat Tracking Results]{Joint Estimation Beat Tracking Results. We observe that for larger datasets, pre-training was actually quite beneficial, and led to a slight increase in beat tracking performance compared to purely supervised, vanilla training. This was not the case for the smaller, Hainsworth \citep{hainsworth2004particle} dataset. The results we obtain in this joint setup are also poorer than those obtained in the pure beat tracking experiment. They also are not up-to-par with the RNN methodology presented in \citep{bock2016joint} (the other metrics were never presented in \citep{bock2016joint}).}
\label{tab:joint_beat}
\end{table}

Overall, for both beat tracking and downbeat estimation, the pure and joint experiments did not indicate that our pre-training method was beneficial for test set performance. Generally, vanilla and pre-trained network performances were within a standard deviation of each other. The training plots in Figures \ref{fig:ball}-\ref{fig:hain}-\ref{fig:gt} (appendix) did however show that pre-training our models led to faster downstream training. This inspired the next experiment, in which we limited the amount of training data our network was exposed to. We then validated and tested it on a constant 25\% of each dataset. In this setting, pre-training was found to be quite beneficial.\footnote{Experiments remain separate for each dataset.}

\begin{table}[H]
\centering
\begin{tabular}{ |p{3cm}|p{1.1cm}|p{1.1cm}|p{1.1cm}|p{1.1cm}|p{1.1cm}|}
\hline
Dataset&\textit{F1}&\textit{CMLc}&\textit{CMLt}&\textit{AMLc}&\textit{AMLt}\\
\hline
\textit{Ballroom \citep{gouyon2006experimental} \citep{krebs2013rhythmic}}& & & & &\\
\textit{- Vanilla}& 0.806 (0.062) & 0.710 (0.107) & 0.712 (0.109) & 0.875 (0.050) & 0.877 (0.051)\\
\textit{- Pre-trained}& 0.822 (0.040) & \textbf{0.767 (0.069)} & \textbf{0.768 (0.069)} & \textbf{0.879 (0.038)} & \textbf{0.881 (0.039)}\\
\textit{- Joint RNN \citep{bock2016joint}} & \textbf{0.863} & NA & NA & NA & NA\\
\hline
\textit{Hainsworth \citep{hainsworth2004particle}}& & & & &\\
\textit{- Vanilla}& 0.517 (0.083) & \textbf{0.452 (0.084)} & \textbf{0.454 (0.083)} & \textbf{0.705 (0.083)} & \textbf{0.712 (0.081)}\\
\textit{- Pre-trained}& 0.501 (0.062) & 0.429 (0.097) & 0.432 (0.097) & 0.692 (0.066) & 0.697 (0.066)\\
\textit{- Joint RNN \citep{bock2016joint}} & \textbf{0.684} & NA & NA & NA & NA\\
\hline
\textit{GTZAN \citep{marchand2015swing} \citep{tzanetakis2002musical}}& & & & &\\
\textit{- Vanilla}& 0.612 (0.043) & 0.527 (0.050) & 0.528 (0.049) & 0.796 (0.039) & 0.799 (0.039)\\
\textit{- Pre-trained}& 0.613 (0.034) & \textbf{0.536 (0.044)} & \textbf{0.537 (0.044)} & \textbf{0.806 (0.031)} & \textbf{0.808 (0.031)}\\
\textit{- Joint RNN \citep{bock2016joint}} & \textbf{0.640} & NA & NA & NA & NA\\
\hline
\end{tabular}
\caption[Joint Estimation Downbeat Estimation Results]{Joint Estimation Downbeat Estimation Results. We observe the same behaviour as we do with beat tracking.}
\label{tab:joint_down}
\end{table}

\paragraph{Impact of Training Set Size on Learning Performance}\mbox{}\\

Our pre-trained model was found to outperform our vanilla model in a limited-data setting. Table \ref{tab:percs} displays our results. As one can see, when the number of training samples is extremely low (i.e. under 10), the pre-trained model significantly outperforms the vanilla model. For beat tracking, the difference is only significant when the number of training samples is very low. In the Ballroom \citep{gouyon2006experimental} \citep{krebs2013rhythmic} dataset, pre-trained and vanilla models average a similar F1-score at around 26 samples, or only 5\% of the training set. The results progress in a similar fashion as the number of training samples increases. On the other hand, when only 1\% of the training set is used (5 samples), the difference in performance is huge. The vanilla model averages an F1-score of 0.281, whereas the pre-trained model averages an F1-score of 0.694. The trend is similar for both the Hainsworth \citep{hainsworth2004particle} and GTZAN \citep{tzanetakis2002musical} \citep{marchand2015swing} datasets. After about 15 to 20 training samples, pre-training seems to become insignificant.

For downbeat estimation, the story is similar. The task does however seem to require a bit more data. In general, model performance is similar after using approximately 10-20\% of the training set. Table \ref{tab:percs} contains all our results for this experiment. Overall, we believe these results to be encouraging. For a few-shot learning task related to musical rhythm, perhaps our pre-trained model initialization could learn more quickly, or even adapt in real-time. It has, after all, shown an ability to learn using very few examples for beat tracking and downbeat estimation tasks.

\begin{table}[H]
\resizebox{\columnwidth}{!}{%
\begin{tabular}{ |p{3.3cm}|p{1cm}|p{1cm}|p{1cm}|p{1cm}|p{1cm}|p{1cm}|p{1cm}| }
\hline
\textit{Dataset}&\textit{1\%}&\textit{2\%}&\textit{5\%}&\textit{10\%}&\textit{20\%}&\textit{50\%}&\textit{75\%} \\
\hline
\hline
\textbf{Ballroom \citep{gouyon2006experimental} \citep{krebs2013rhythmic}}& & & & & & & \\
\hline
- Train Set Size& 5 & 10 & 26 & 51 & 103 & 257 & 386 \\
\hline
- Vanilla Beat F1& 0.281 (0.008) & 0.677 (0.098) & \textbf{0.739 (0.100)} & \textbf{0.786 (0.045)} & \textbf{0.860 (0.014)} & \textbf{0.907 (0.018)} & \textbf{0.927 (0.021)} \\
- Pretrain Beat F1& \textbf{0.694 (0.038)} & \textbf{0.737 (0.027)} & 0.727 (0.066) & 0.775 (0.060) & 0.825 (0.025) & 0.877 (0.020) & 0.908 (0.015) \\
\hline
- Vanilla Down F1& 0.061 (0.007) & 0.353 (0.148) & \textbf{0.536 (0.049)} & 0.576 (0.050) & \textbf{0.701 (0.015)} & \textbf{0.776 (0.038)} & \textbf{0.839 (0.028)} \\
- Pretrain Down F1& \textbf{0.423 (0.045)} & \textbf{0.476 (0.038)} & 0.506 (0.074) & \textbf{0.579 (0.075)} & 0.644 (0.052) & 0.741 (0.052) & 0.802 (0.031) \\
\hline
\hline
\textbf{Hainsworth \citep{hainsworth2004particle}}& & & & & & & \\
\hline
- Train Set Size& 2 & 3 & 8 & 17 & 33 & 83 & 125 \\
\hline
- Vanilla Beat F1& 0.273 (0.013) & 0.278 (0.016) & 0.437 (0.141) & 0.606 (0.033) & \textbf{0.631 (0.014)} & 0.655 (0.093) & \textbf{0.717 (0.020)} \\
- Pretrain Beat F1& \textbf{0.489 (0.054)} & \textbf{0.498 (0.113)} & \textbf{0.588 (0.040)} & \textbf{0.612 (0.025)} & 0.597 (0.032) & \textbf{0.695 (0.020)} & 0.708 (0.026) \\
\hline
- Vanilla Down F1& 0.063 (0.009) & 0.067 (0.005) & 0.074 (0.036) & 0.226 (0.078) & 0.279 (0.072) & 0.376 (0.075) & \textbf{0.462 (0.038)} \\
- Pretrain Down F1& \textbf{0.167 (0.078)} & \textbf{0.208 (0.075)} & \textbf{0.245 (0.049)} & \textbf{0.303 (0.056)} & \textbf{0.307 (0.082)} & \textbf{0.389 (0.065)} & 0.434 (0.056) \\
\hline
\hline
\textbf{GTZAN \citep{marchand2015swing} \citep{tzanetakis2002musical}}& & & & & & & \\
\hline
- Train Set Size& 8 & 15 & 38 & 75 & 150 & 375 & 563 \\
\hline
- Vanilla Beat F1& 0.495 (0.125) & \textbf{0.739 (0.024)} & \textbf{0.803 (0.016)} & \textbf{0.784 (0.027)} & \textbf{0.814 (0.011)} & 0.811 (0.010) & 0.819 (0.010) \\
- Pretrain Beat F1& \textbf{0.656 (0.057)} & 0.701 (0.054) & 0.741 (0.040) & 0.783 (0.020) & 0.803 (0.019) & \textbf{0.820 (0.009)} & \textbf{0.831 (0.015)} \\
\hline
- Vanilla Down F1& 0.089 (0.089) & 0.135 (0.022) & \textbf{0.430 (0.026)} & 0.470 (0.044) & 0.506 (0.026) & 0.537 (0.027) & \textbf{0.580 (0.016)} \\
- Pretrain Down F1& \textbf{0.325 (0.048)} & \textbf{0.380 (0.043)} & 0.424 (0.044) & \textbf{0.482 (0.049)} & \textbf{0.515 (0.033)} & \textbf{0.572 (0.030)} & 0.573 (0.019) \\
\hline
\end{tabular}%
}
\caption[Impact of Training Set Size on Learning Performance]{Impact of Training Set Size on Learning Performance. For each training set percentage, we randomly select a subset of the training set for model training. The validation and test sets are constant for each dataset (each 12.5\% of a the dataset). We report the mean and standard deviation of 10 experiments for each percentage.}
\label{tab:percs}
\end{table}

\begin{table}[H]
\centering
\begin{tabular}{|p{3.3cm}|p{1cm}|p{1cm}|p{1cm}|p{1cm}|p{1cm}|}
\hline
\textit{Training Dataset}&\textit{F1}&\textit{CMLc}&\textit{CMLt}&\textit{AMLc}&\textit{AMLt}\\
\hline
\hline
\textbf{Ballroom \citep{gouyon2006experimental} \citep{krebs2013rhythmic}}& & & & &\\
\hline
- Vanilla Beat F1 & \textbf{0.715 (0.013)} & \textbf{0.475 (0.021)} & \textbf{0.569 (0.024)} & \textbf{0.616 (0.017)} & \textbf{0.760 (0.017)} \\
- Pretrain Beat F1 & 0.699 (0.011) & 0.447 (0.019) & 0.545 (0.022) & 0.571 (0.011) & 0.714 (0.016) \\
\hline
- Vanilla Down F1& 0.478 (0.025) & 0.403 (0.028) & 0.412 (0.026) & 0.641 (0.018) & 0.656 (0.017)\\
- Pretrain Down F1& 0.468 (0.015) & 0.391 (0.023) & 0.401 (0.025) & 0.619 (0.017) & 0.641 (0.017)\\
\hline
\hline
\textbf{GTZAN \citep{marchand2015swing} \citep{tzanetakis2002musical}}& & & & &\\
\hline
- Vanilla Beat F1& \textbf{0.759 (0.014)} & 0.568 (0.030) & \textbf{0.690 (0.030)} & 0.672 (0.018) & 0.820 (0.011)\\
- Pretrain Beat F1& 0.756 (0.006) & \textbf{0.576 (0.016)} & 0.682 (0.011) & \textbf{0.695 (0.019)} & \textbf{0.830 (0.010)}\\
\hline
- Vanilla Down F1& 0.529 (0.012) & 0.494 (0.017) & 0.501 (0.017) & 0.730 (0.016) & 0.741 (0.016)\\
- Pretrain Down F1& \textbf{0.534 (0.013)} & \textbf{0.505 (0.011)} & \textbf{0.509 (0.012)} & \textbf{0.734 (0.020)} & \textbf{0.742 (0.019)}\\
\hline
\end{tabular}
\caption[Hainsworth \citep{hainsworth2004particle} Mean Test Set Results]{Hainsworth \citep{hainsworth2004particle} Mean Test Set Results. Each model was trained on $\frac{7}{8}$ths of the training data set and validated on the last fold. We report the mean and standard deviation for each evaluation metric on beat tracking and downbeat estimation. The experiment was conducted in a joint estimation setting using 8-fold CV on each train set.}
\label{tab:cross}
\end{table}

\paragraph{Cross-data Set Generalization}\mbox{}\\

Finally, we did not notice that pre-training aided our model to better generalize to new data. When being trained on the Ballroom \citep{gouyon2006experimental} \citep{krebs2013rhythmic} data set, our pre-trained model performed worse than our vanilla model on the Hainsworth data set \citep{hainsworth2004particle}. For beat tracking, the vanilla model averaged an F1-score of 0.715 whereas the pre-trained model averaged a score of 0.699 (for downbeat estimation, we obtain scores of 0.478 and 0.468). For the GTZAN \citep{gouyon2006experimental} \citep{krebs2013rhythmic} dataset, we found the pre-trained model to perform slightly better than the vanilla one. In both cases, however, the evaluation metrics are quite similar. Table \ref{tab:cross} displays these results. This is most likely due to our pre-trained model overfitting to the training set once again. 
\chapter{Discussion and Future Work}\label{ch:mathtest} 



Although our network learned an onset function that resembles a musical beat activation function, the performance we obtained after pre-training our network was quite disappointing. Pre-training our network was never found to elevate the performance on our downstream tasks drastically. The representations that were learned were also unable to be used in a frozen setting.

In hindsight, the data we trained our pretext task on may have been too limited. Perhaps using "badly" separated tracks would have helped our model learn better representations. This would have enabled us to safely separate our downstream audio into ROS and drum stems.\footnote{Separating our downstream songs is risky with the network trained as is. Numerous tracks do not contain drums. The stem RMS thresholding severely biases our model towards learning a drum audio representation.} By doing so, the model's transfer learning would be more natural; both pretext and downstream tasks would be source separated. 

The onset function computed by the pre-trained model could then have served as input to a new, small-sized model which predicts beats and downbeats. We could also have compared our anchors to each other throughout each batch in contrastive learning. Doing so would have made the pretext task harder. On the flip side, distinguishing drum-less sounds from each other would have enabled our model to learn better representations for ROS stems, which are currently less well-defined and onset-like than drum stem output vectors.

There is however a silver lining. For starters, we believe that the onset function output by the model must be investigated more thoroughly. Are the periodicities observed interpretable? Can one just shift them by a certain value to obtain beat annotations? Or could they be used as is for automatic tempo estimation? We hope to address these questions in the upcoming weeks by first relaxing the RMS constraint present in our work, re-training our network on more data, and finally creating a network which turns the model's onset functions into beat annotations. Perhaps using networks that specialize in representing different stems for both the pretext and downstream tasks could aid with performance (i.e. one model specializes in representing drum stems, whilst the other specializes in representing ROS stems in the pretext task; both are then fine-tuned for the downstream task).\footnote{Note that we tried creating separate networks for drum and ROS stems for the pretext task. We struggled to train the model however.}

More generally, we also believe that our work can serve as inspiration for numerous other experiments. In the realm of contrastive learning, we truly believe that audio source separation algorithms such as Spleeter \citep{spleeter2020} could serve as a data augmentation technique tailored to musical representation learning. A key property of music is that it mixes sounds from one or more sources to produce a whole. By deconstructing this whole randomly using source separation, a model would be able to understand how these sources interact with respect to each other, and by extension understand this key property of music.

Within contrastive learning, we also believe that audio source separation could be used to tailor our training towards a particular embedding we would want to learn. Maybe we could extract a large number of vocals and train a model to match snippets that come from the same song? This could potentially be a good pretext task for a downstream task such as singer identification, which is quite popular in MIR. Another option would be to match snippets of Spleeter-generated acapella versions of a song (anchors) to full song mixtures (positives). This would allow a model to better recognize vocals embedded within other instruments.

These are all open questions. In the immediate future, we hope to address those that pertain to our pretext and downstream tasks. The rest are open for anyone to try.

\chapter{Conclusion}\label{ch:mathtest} 

In this thesis, we present a new self-supervised learning pretext task targeted to beat tracking downstream tasks. This pretext task is centred around the idea of matching drum stems to their appropriate ROS signal. A ROS stem is a track without its drums. In order to achieve this task, we made sure both the ROS and drum stems had similar RMS values. The VQT representations of each stem were then used as input to a fully-convolutional and recurrent model that was to solve a contrastive learning task. The ROS stems became the anchors, while the drum stems became the positives (and by extension the negatives). The resulting model was found to output an onset function, where "spikes" are representative of musical events. These events are often repetitive, suggesting that, with a bit of refining, our approach can be adapted for beat tracking. 

When fine-tuning our model to popular beat tracking and downbeat estimation datasets, we found that our pre-trained model was unable to consistently surpass its vanilla counterpart. The only setting in which it was clearly able to was when the training set was limited data-wise (usually when less than 5\% of the train set was used). Our pre-training process also did not help our model generalize to other datasets. In the future, we hope to come back to some of the ideas introduced in this thesis. We believe source separation can become a core component of the data augmentation process in musical contrastive learning, and hope to see more work exploring this idea in the future.

\appendix
\chapter{Appendix}

\section{Chapter 6}

\subsection{Downstream Tasks}

\begin{figure}[H]
\centering
\subfloat[][Beat Tracking (Random Initialization)]{\includegraphics[width=.5\textwidth]{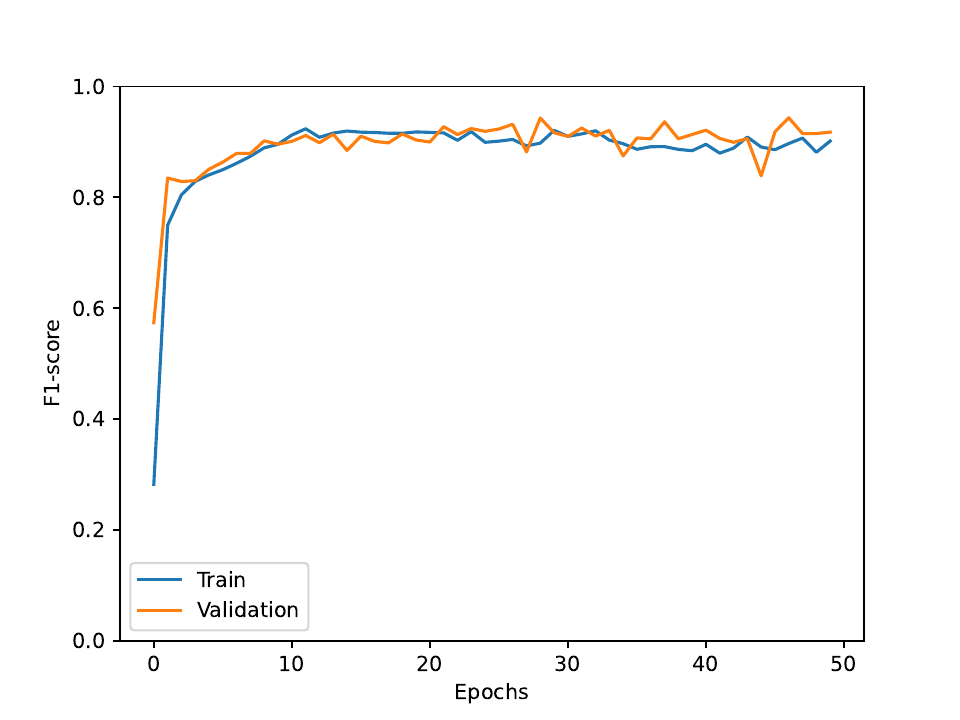}}
\subfloat[][Beat Tracking (Pre-Trained Initialization)]{\includegraphics[width=.5\textwidth]{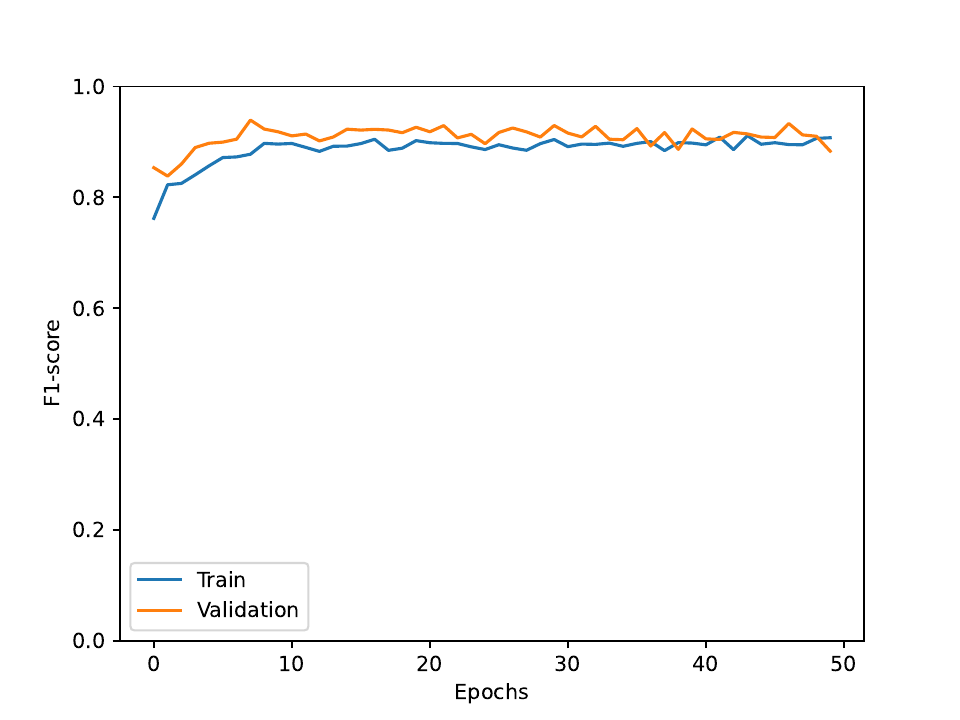}}\\
\subfloat[][Downbeat Estimation (Random Initialization)]{\includegraphics[width=.5\textwidth]{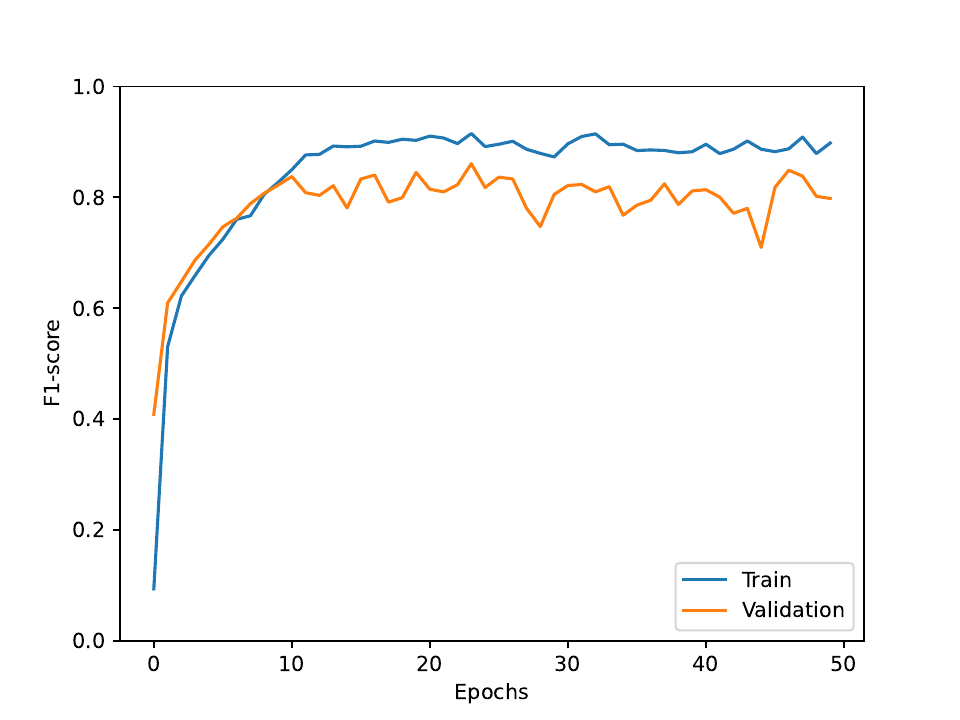}}
\subfloat[][Downbeat Estimation (Pre-Trained Initialization)]{\includegraphics[width=.5\textwidth]{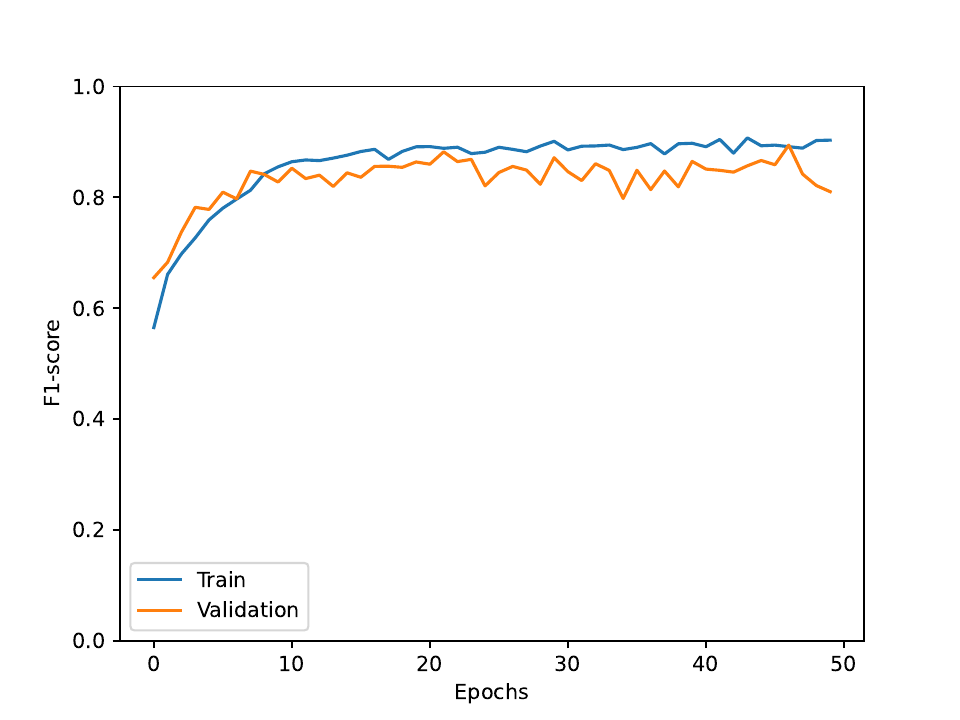}}
\caption{Ballroom \citep{gouyon2006experimental} \citep{krebs2013rhythmic} F1-score on Train and Validation Sets}
\label{fig:ball}
\end{figure}

\begin{figure}[H]
\centering
\subfloat[][Beat Tracking (Random Initialization)]{\includegraphics[width=.5\textwidth]{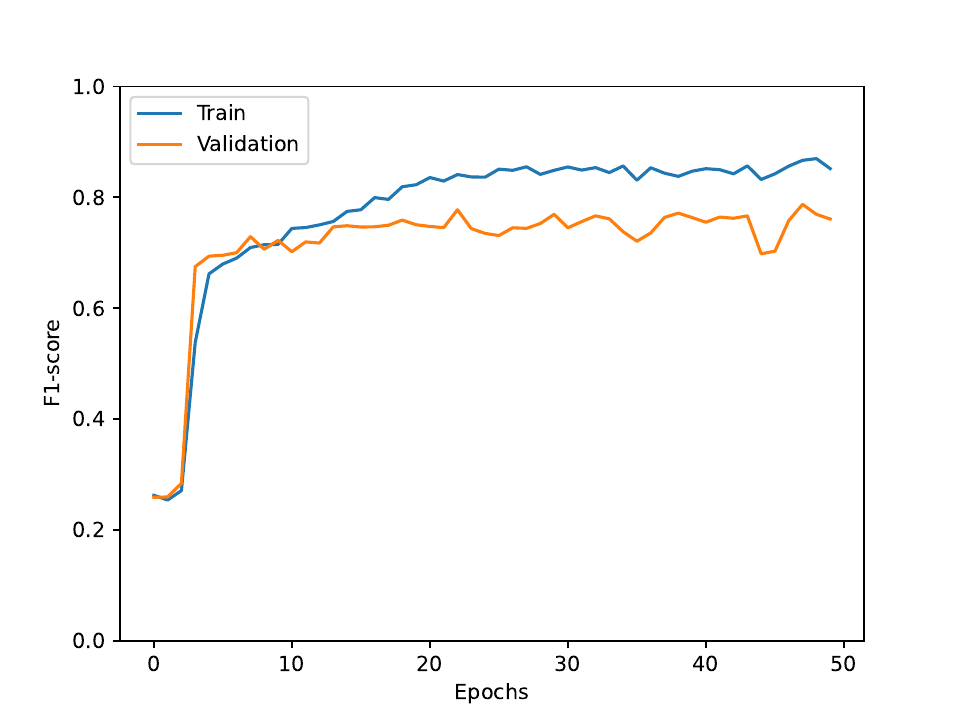}}
\subfloat[][Beat Tracking (Pre-Trained Initialization)]{\includegraphics[width=.5\textwidth]{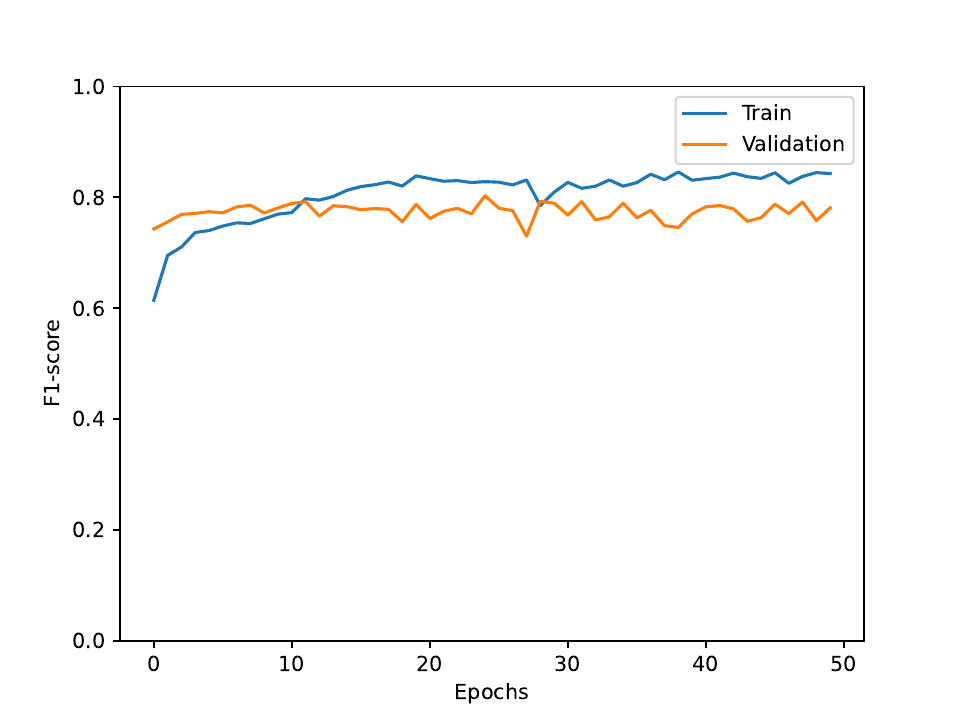}}\\
\subfloat[][Downbeat Estimation (Random Initialization)]{\includegraphics[width=.5\textwidth]{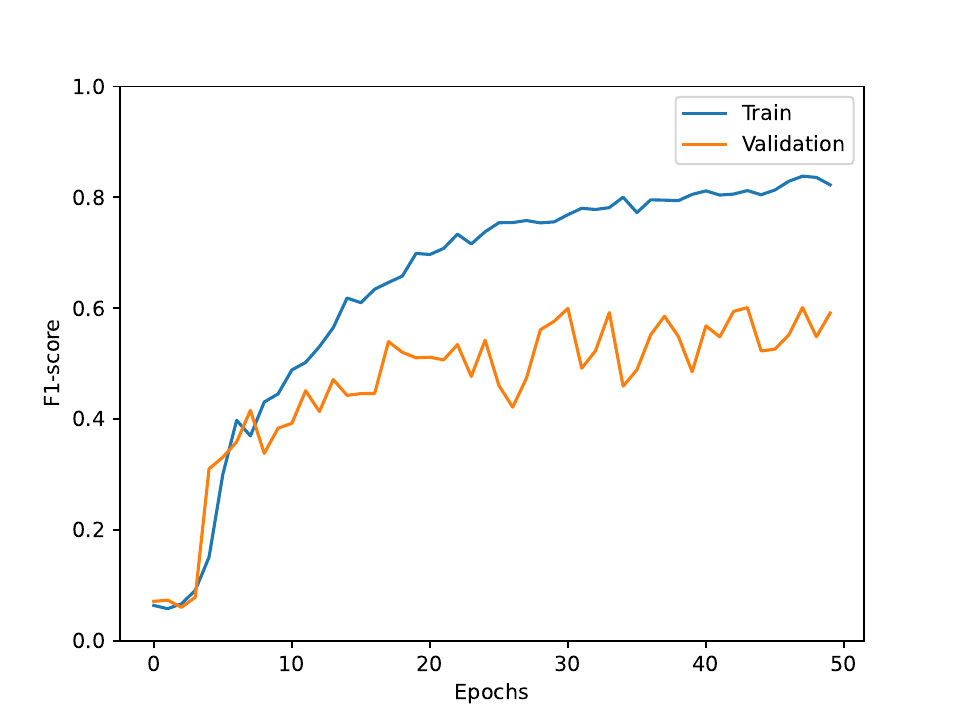}}
\subfloat[][Downbeat Estimation (Pre-Trained Initialization)]{\includegraphics[width=.5\textwidth]{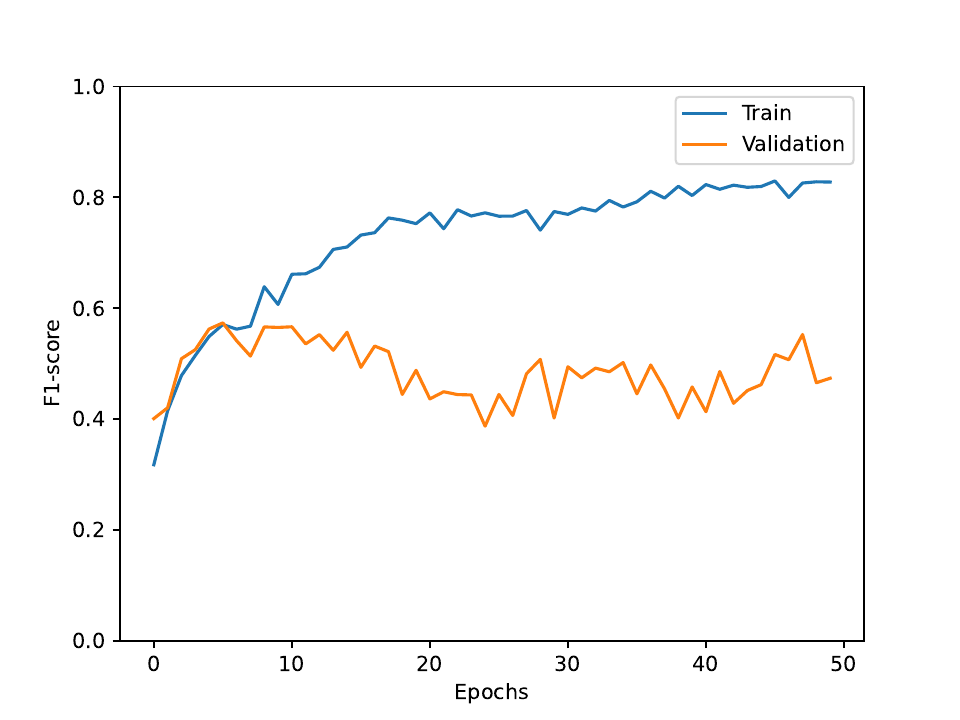}}
\caption{Hainsworth \citep{hainsworth2004particle} F1-score on Train and Validation Sets}
\label{fig:hain}
\end{figure}

\begin{figure}[H]
\centering
\subfloat[][Beat Tracking (Random Initialization)]{\includegraphics[width=.5\textwidth]{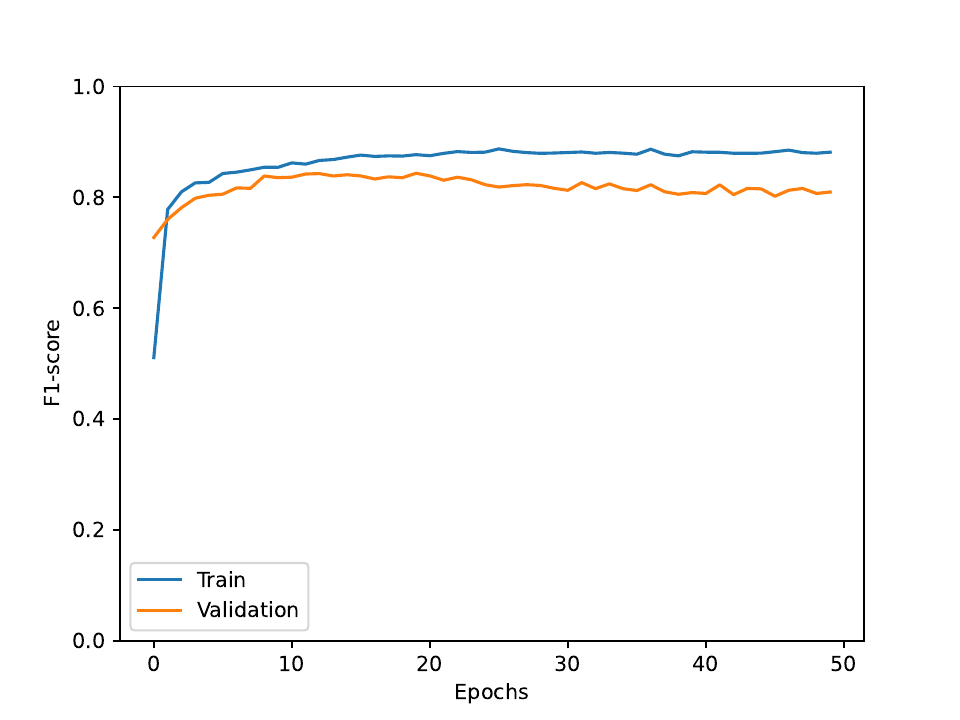}}
\subfloat[][Beat Tracking (Pre-Trained Initialization)]{\includegraphics[width=.5\textwidth]{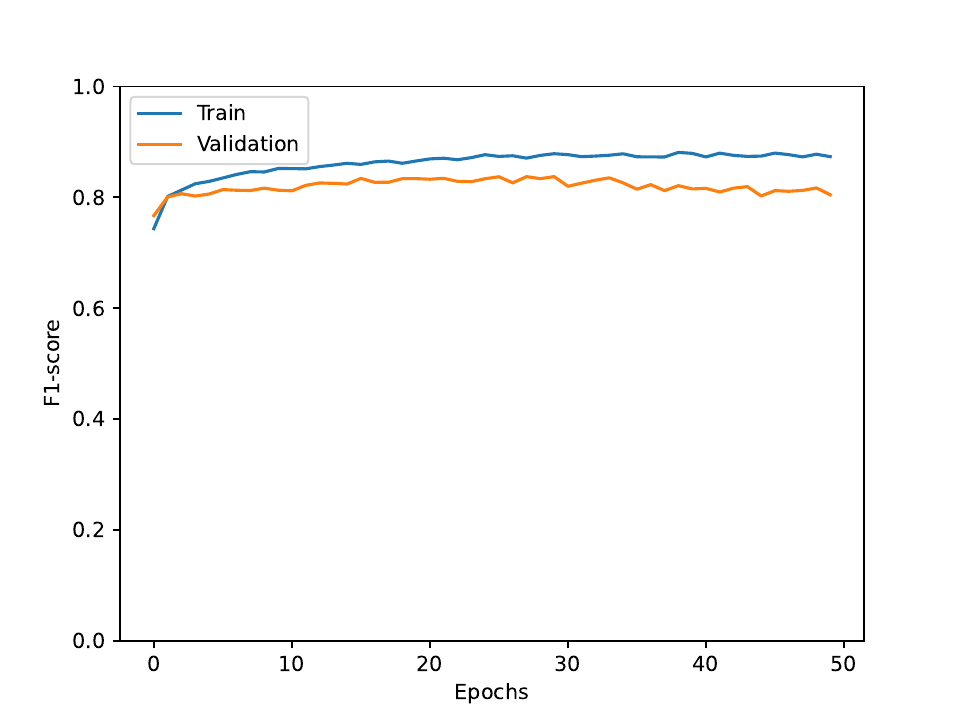}}\\
\subfloat[][Downbeat Estimation (Random Initialization)]{\includegraphics[width=.5\textwidth]{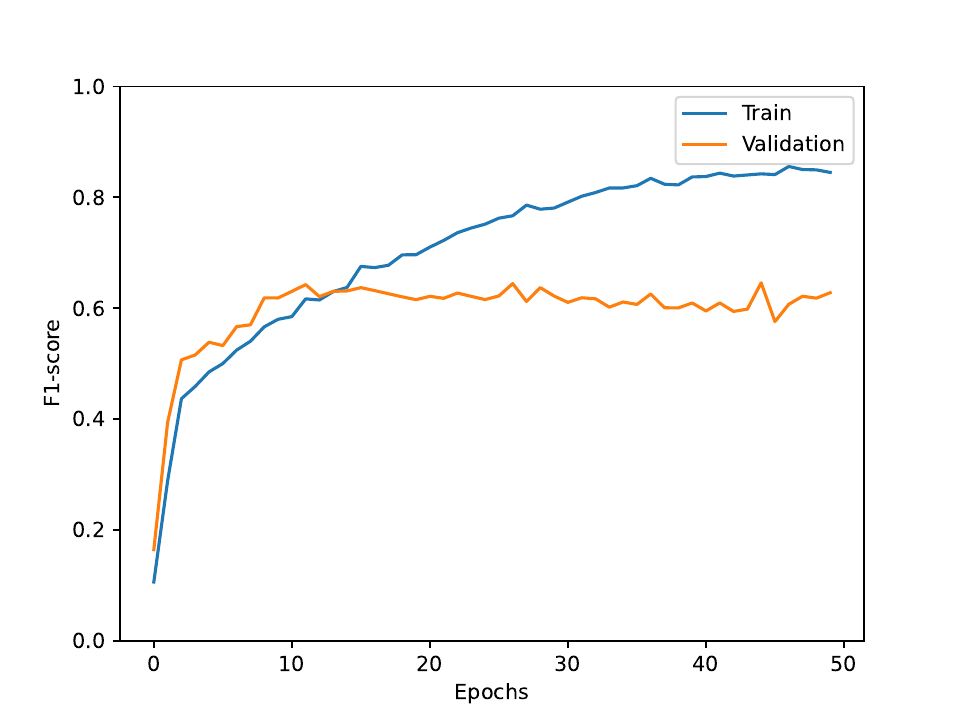}}
\subfloat[][Downbeat Estimation (Pre-Trained Initialization)]{\includegraphics[width=.5\textwidth]{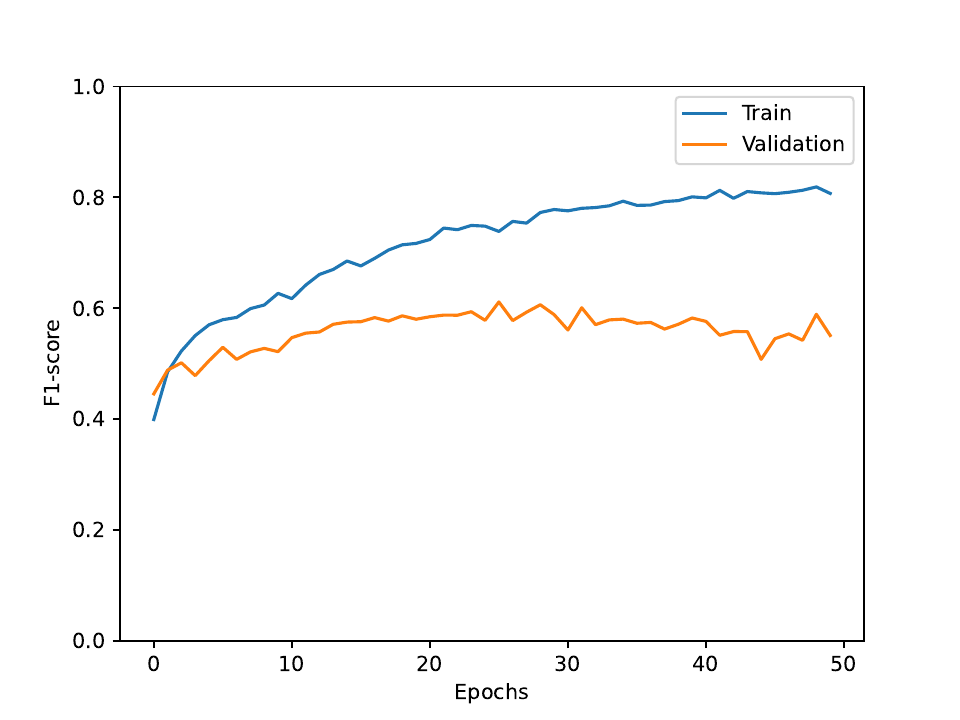}}
\caption{GTZAN \citep{tzanetakis2002musical} \citep{marchand2015swing} F1-score on Train and Validation Sets}
\label{fig:gt}
\end{figure}
\manualmark
\markboth{\spacedlowsmallcaps{\bibname}}{\spacedlowsmallcaps{\bibname}} 
\refstepcounter{dummy}
\addtocontents{toc}{\protect\vspace{\beforebibskip}} 
\addcontentsline{toc}{chapter}{\tocEntry{\bibname}}
\label{app:bibliography}
\printbibliography

\end{document}